\documentclass[a4paper]{article}
\pdfoutput=1
\usepackage{amsmath}
\usepackage{amssymb}
\usepackage{layout}
\usepackage{ifthen}
\usepackage{url}
\usepackage{pgf}
\usepackage{tikz}
\usepackage[retainorgcmds]{IEEEtrantools}
\usepackage{lscape}
\usetikzlibrary{calc,intersections,through,backgrounds}
\author{Roger Sewell\\
\href{mailto:roger.sewell@cantab.net}{\scriptsize{roger.sewell@cantab.net}}
}
\usepackage{graphicx}
\usepackage{xr}
\usepackage{float}
\usepackage[pdftex,colorlinks=true,linkcolor=blue]{hyperref}

\newboolean{isfalse}
\setboolean{isfalse}{false}

\title{Simulation and inference on purely observational
  methods of monitoring vaccine effectiveness post-deployment: none is
  reliable without precise information on population
  behaviour\footnote{Fourth version deposited in arxiv; RFS version 1.41.1.1
; added conditional independence relationships and a third way to
  resolve the problem.}
}

\begin{document}
\maketitle

Acknowledgements: I am grateful to Matt Jackson for pointing out that
UK media don't all prefer the TNCC method, for detailed comments on
the abstract, and for suggesting applying the model to partial
real-world data; and to Stephen Mack for reviewing the main text.


\begin{center}
\textbf{Abstract}
\end{center}

Two observational methods are currently being used to monitor
post-deployment vaccine effectiveness against infection: the obvious
crude method comparing rate of testing positive for infection per head
of vaccinated population with that rate per head of unvaccinated
population; and the test-negative case control (TNCC) method. The two
methods give very different results. Various parties' preference for
choice of method appears to broadly coincide with their vested
interests in getting the result that method gives. We want to know
whether either method is reliable.

We assume either a homogeneous population or one partitioned into two
homogeneous subsets which differ only in their not-directly-observable
healthcare-seeking behaviour including probability of getting
vaccinated. We first consider uniform independent priors on the
probabilities of being hospitalised conditional on subset, vaccination
status, and infection status. We simulate from the resulting model and
observe the TNCC estimate, the crude estimate, and the Bayesian
central 95\% confidence interval on vaccine effectiveness represented
as log ratio of odds ratios for infection with and without
vaccination. To show that such partial real-world data as is available
leads to similar conclusions we also apply inference with the model to
an example of real world data from 2021.

With these wide open priors, even when the population is homogeneous,
the Bayesian 95\% confidence interval typically has a width of nearly
4 nats (55-fold), implying too much uncertainty in the inference for
the data currently collected to be of any use in monitoring vaccine
effectiveness. While we do find that there exist some tight priors
under which the data is useful, some lead to TNCC being more accurate
(by around 5-fold based on standard deviation in the log domain) and
with others the crude estimate is more accurate (by around 5 fold
likewise).

Thus using only data from those spontaneously choosing to be tested,
we find that neither TNCC nor the crude method is reliably better than
the other, and indeed that the desired information is not present in
this data. We conclude that effective post-deployment monitoring of
vaccine effectiveness and side-effects requires either (a) strong
information on the population’s healthcare-seeking behaviour, (b) only
considering tests administered to a random sample of the population
independent of their vaccination status and perceived state of health,
or (c) ongoing randomised controlled trials (RCTs), rather than just
choosing whichever of TNCC and crude estimate gives the result we
prefer to find. We favour using RCTs.

\tableofcontents

\section{Introduction}

When doing post-marketing monitoring of efficacy of vaccines
unexpected results are often obtained by collecting the obvious data,
and the Test-Negative Case Control method (TNCC) is often put forward
\cite{Chua, KirsebomLancet} as a suitable method of ensuring that the
biases thought to be responsible for these results are removed and the
underlying vaccine efficacy made clear.

To take an example that is topical at the time of writing, in table 11
of \cite{Week48} the United Kingdom Health Security Agency (UKHSA)
reports raw data suggesting that the apparent rate of covid-19
infection in vaccinated 50-59 year olds is approximately double that
in the corresponding unvaccinated population. In figure 1 of the same
report they report data based on the TNCC method suggesting that, when
the relevant biases are removed, the infection rate is between 45\%
and 95\% lower in the vaccinated population than the unvaccinated. Or
in another example, in the over-65s, \cite{Stowe} reports vaccine
effectiveness (VE) values between 85\% and 93\% by TNCC while
observing the raw rate of hospital admission for covid disease in the
vaccinated to be 51\% of that in unvaccinated individuals (and indeed
the rate for vaccinated individuals for non-covid disease to be five times
higher than for unvaccinated individuals)\footnote{Based on ~10
  million vaccinated and 600 thousand unvaccinated in the over-65s in
  the UK at the time, and the data in Table S12 of \cite{Stowe}.}.

The TNCC method has been used since at least 1980 \cite{Broome}, and
for monitoring influenza vaccine effectiveness in particular since at
least 2005, and has been reviewed extensively in that context in
\cite{Sullivan} in 2016. As these latter authors make clear through
the use of Bayesian networks represented by directed acyclic graphs
(DAGs), it compensates perfectly for a situation where the population
is divided into two groups, one of which always seeks healthcare when
ill, while the other never seeks healthcare when ill. Where other
covariates are measured, the combination of TNCC with logistic
regression allows at least partial compensation for other variation
within the population, e.g. age, comorbidity, etc. The same authors
point out, as agreed in \cite{Westreich}, that where the source of
bias is not so binary and cut-and-dried there is no guarantee that
bias due to variations in healthcare-seeking behaviour will be
correctly handled.

Therefore the purpose of this paper is to investigate whether the TNCC
method is more or less reliable than the crude method of estimating
vaccine effectiveness in such a post-marketing situation, and indeed
whether either method is at all suitable for this purpose. Even in the
simple situation where the population is binarily divided into two
groups, each of which has their own probabilities of seeking
healthcare when ill with or without covid infection, we will suggest
that neither method is adequate for the purpose, neither is free of
bias, and that in fact the observational data itself does not contain
the information sought to any useful degree of accuracy. We therefore
suggest that in a post-marketing situation a randomly chosen subset of
the population should be offered ongoing participation in a
prospective randomised controlled study of vaccine versus placebo,
without excluding any part of the population in which the vaccine is
actually being used.

We first define intuitively the TNCC and the crude methods of
observationally measuring vaccine effectiveness (VE), and how our
model works (precise definitions and descriptions are given in the
appendices). We describe the results under various different priors,
then apply inference with the model to partial real-world data from
\cite{LopezBernalNEJM}, before discussing the implications and the
actions we believe to be necessary.

Throughout we will refer to the disease being vaccinated against as
``covid'', though there is nothing in the work reported other than the
last-mentioned application to the data from \cite{LopezBernalNEJM}
that is in any way specific to this particular disease.

\section{Intuitive description of methods}

Full details of all definitions and methods are in appendices
\ref{defs} and \ref{model}. Here we attempt to give an intuitive
description of what has been done.

\subsection{Measure of vaccine effectiveness}

The usual definition of vaccine effectiveness is $$E = 1 -
\frac{P(L|V)}{P(L|v)},$$ where $L$ denotes the event that a patient
has covid, $V$ the event that they are vaccinated, $v$ the event that
they are not vaccinated, and $P(A|B)$ denotes the probability of $A$
given $B$. In this paper we instead use the measure $$T_0 =
\log\left(\frac{P(L|V)}{1 - P(L|V)} / \frac{P(L|v)}{1 -
  P(L|v)}\right),$$ the log ratio of the odds of getting covid when
vaccinated to the odds when not vaccinated. If the fraction of the
population with covid is small then $T_0$ is approximately the same as
$\log(1-E)$; however, unlike $\log(1-E)$, $T_0$ also notices when the
probability of \textit{not} getting covid is increased from a small
value to a slightly larger one: see the examples in table
\ref{T0examples}.

\begin{table}[hpt]
\begin{center}
\begin{tabular}{c|c|c|c}
$P(L|v)$ & $P(L|V)$ & $\log(1-E)$ & $T_0$\\ \hline
$0.5$ & $0.5$ & $0$ & $0$\\ \hline
$0.002$ & $0.001$ & $-0.693$ & $-0.694$ \\ \hline
$0.999$ & $0.998$ & $-0.001$ & $-0.694$ \\ \hline
$0.2$ & $0.1$ & $-0.693$ & $-0.811$\\ \hline
$0.9$ & $0.8$ & $-0.118$ & $-0.811$\\ \hline
$0.1$ & $0.2$ & $+0.693$ & $+0.811$\\ \hline
\end{tabular}
\caption{Examples of the relationship between the probabilities of
  getting covid with and without vaccination and the corresponding
  values of $E$ and $T_0$.}
\label{T0examples}
\end{center}
\end{table}

In either case, a negative value of $\log(1-E)$ or $T_0$ indicates the
vaccine is effective, and a positive value indicates that the vaccine
makes the recipient \textit{more} likely to get covid than if he
hadn't been vaccinated. The reasons for the choice of $T_0$ rather
than $E$ are that it makes the values returned by TNCC and/or the
crude estimate correspond to $T_0$ in a manner that is easy to display
and check without many samples accruing near zero or one and being
hard to distinguish, and that it makes an increase in an initially
small probability of avoiding infection easier to
detect\footnote{While this may not be important in the case of covid,
  it may be very important for an always-fatal infection that one is
  very likely to get.}.

\subsection{Estimates of vaccine effectiveness}

$T_1$ will denote the corresponding estimate of vaccine effectiveness
returned by the TNCC method; this is estimated by comparing the odds
of having been vaccinated in those hospitalised with covid with the
odds of having been vaccinated in those hospitalised without covid. In
an idealised situation where everybody is always hospitalised for any
illness $T_1$ will be very nearly equal to $T_0$; this will also be
true in situations when the population divides into two subsets, one
of which is always hospitalised for any illness and the other never
\cite{Sullivan}. Further, $T_1=T_0$ if hospitalisation is
conditionally independent of vaccination status given infection
status, or if hospitalisation is conditionally independent of
infection status given vaccination status -- but neither of these is
remotely likely to be true.

$T_2$ will denote the corresponding crude estimate of vaccine
effectiveness that results from assuming that dividing the number of
unvaccinated hospitalised covid cases by the number of unvaccinated
people in the population gives an unbiased estimate of the infection
rate in the unvaccinated (and similarly for the vaccinated). Again, if
everybody is always hospitalised for every illness, $T_2$ will be very
nearly equal to $T_0$; but in this case this will not be true if the
population divides into two subsets with different healthcare-seeking
behaviours -- which is the reason why TNCC is often used. Moreover, if
hospitalisation is conditionally independent of vaccination status
given infection status, then again we will have $T_2=T_0$.

What we are interested in is whether $T_1$ or $T_2$ is a better
estimator of $T_0$, and in whether the better of the two is a good
enough estimator for practical use.

\subsection{Model}

Our model will divide the population into two homogeneous subsets (0
and 1) differing only in the non-directly-observable characteristic of
the nature of their healthcare-seeking behaviour. The fraction of the
population in subset 1 is given by $p$, an unknown on which we put a
prior distribution (in the base case a uniform distribution on
$[0,1]$).

Each patient is then vaccinated with a probability $r_s$ that depends
on which subset $s$ the patient is in; in the base case these two
probabilities are again distributed uniformly and independently on
$[0,1]$. $v$ will be 1 if the patient is vaccinated and 0 otherwise.

Each patient then does or doesn't acquire covid, with the probability
of getting covid being $j_v$, depending on whether the patient has
been vaccinated; then the variable $l$ will be 1 if the patient gets
covid, otherwise 0. Obviously one hopes that $j_1 < j_0$, but we don't
assume this as we don't know whether this vaccine protects against
covid or does the opposite (and we remind the reader that here
``covid'' means whatever disease the vaccine has been developed
for). In the base case we therefore again put uniform independent
priors on $j_0$ and $j_1$.

Finally each patient does or doesn't get hospitalised ($h=1$ or $h=0$
respectively), with the probability of being hospitalised being
$q_{s,l,v}$ and depending on which subset of the population the
patient is in, whether or not they have covid, and whether or not they
have been vaccinated. To keep our model simple we assume that a
patient is tested for infection if and only if they get
hospitalised. In particular $q_{s,0,1}$ denotes the probability that a
patient in subset $s$ who has been vaccinated and does not have covid
gets hospitalised -- whether from a vaccine induced injury or from
intercurrent unrelated disease. In the base case we again put uniform
independent priors on each of these eight variables $q_{s,l,v}$.

\subsection{Priors}

As noted above, in the base case all the 13 variables constituting
$p,r,j,q$ are given independent uniform priors, so that we can see
what happens when these probabilities are all unknown; we name this
\textbf{Prior 0}, or \textbf{``wide open''}. We will also consider
three other priors, which we name Priors 1, 2, and 3.

The full details of these priors are given in appendix section
\ref{priordescriptions}. In both the base case and the variants the
priors on $j_v$ are both uniform on $[0,1]$. In other respects the
variants differ in ways whose intuitive descriptions are as follows:

\begin{description}

\item[Prior 1:] The population divides into two roughly equal sized
  subsets. One subset mostly gets vaccinated, the other mostly
  not. The mostly-vaccinated subset almost always get hospitalised for
  any illness, while the mostly-unvaccinated subset only get
  hospitalised with probability 0.4 (independent of whether or not
  their illness is covid).

\item[Prior 2:] More than 99\% of the population are in the larger
  of the two subsets; otherwise their behaviour is as in the base
  case.

\item[Prior 3:] Again more than 99\% of the population are in the
  larger of the two subsets, and in both subsets individuals are
  vaccinated with probability 0.5. In either subset, if they either
  have covid or have been vaccinated they all almost always get
  hospitalised for any illness, but otherwise they are hospitalised
  with probability only 0.1 . (They might for example be scared by the
  fact that they have covid, or worried that they have a vaccine
  side-effect if they don't have covid but have been vaccinated.)

\end{description}

\subsection{Exploration of the model}

For each experiment we draw 20 random samples from the model. Each
such sample includes values for each of the probabilities comprising
$p,r,j,q$, and for each of the $N$ patients draws values based on
those probabilities for $s,v,l,h$. We then observe $v$ and $h$ for
every patient, and $l$ just for those patients who are hospitalised
(who get tested for covid).

From these observations we can work out the estimates $T_1$ and $T_2$
of $T_0$. From the value of $j$ we can work out the true value
$T_{0,\text{true}}$ of $T_0$.

With far more effort we can draw samples of the Bayesian posterior
distribution of $T_0$ using the methods described in appendix section
\ref{MCMC}, which give us a picture of the possible range within which
the true value of $T_0$ might lie. In particular we can determine the
2.5 and 97.5 centiles of that distribution, giving us a Bayesian
confidence interval (a.k.a. credible interval) for $T_0$, which is an
interval in which, given the observed data, we are 95\% sure that
$T_0$ lies. 

For the prior distribution in use for this experiment it is not
possible for any other signal processing or analysis technique to
improve on this posterior distribution. Therefore if the difference
between the 2.5 and 97.5 centiles is (say) 4 nats\footnote{A ``nat''
  is a difference of 1 for a natural logarithm to base $e$, and
  represents an $e$-fold difference in the quantity whose logarithm is
  being considered.} then we are uncertain of the value of $e^{T_0}$,
the ratio of odds ratios, to within a factor of $e^4\approx 55$-fold.

The relationships between $P(L|v)$, $T_0$, and $E$ are indicated in
table \ref{Etable}.

\begin{table}[hpt]
\begin{center}
\begin{tabular}{c|c|c|c|c|c|c|c|c|c}
& \multicolumn{9}{c}{$T_0$}\\ \cline{2-10}
$P(L|v)=j_0$ & $-6$ & $-4$ & $-2$ & $-1$ & $0$ & $1$ & $2$ & $4$ & $6$ \\\hline 
$0.01$& $99.7$ & $98.2$ & $86.3$ & $63.0$ & $0.0$ & $-167.2$ & $-594.5$ & $-3454.6$ & $-7929.6$ \\\hline 
$0.05$& $99.7$ & $98.1$ & $85.9$ & $62.0$ & $-0.0$ & $-150.3$ & $-460.0$ & $-1383.7$ & $-1810.0$ \\\hline 
$0.50$& $99.5$ & $96.4$ & $76.2$ & $46.2$ & $0.0$ & $-46.2$ & $-76.2$ & $-96.4$ & $-99.5$ \\\hline 
$0.95$& $95.3$ & $72.8$ & $24.2$ & $7.9$ & $0.0$ & $-3.3$ & $-4.5$ & $-5.2$ & $-5.2$ \\\hline 
$0.99$& $80.1$ & $34.9$ & $6.0$ & $1.7$ & $0.0$ & $-0.6$ & $-0.9$ & $-1.0$ & $-1.0$ \\\hline 
\end{tabular}
\caption{Percentage values of vaccine effectiveness $E$ corresponding
  to various combinations of $P(L|v)$ and $T_0$.
\label{Etable}
}
\end{center}
\end{table}

For $j_0=0.5$ (half of unvaccinated patients get covid), for example,
a difference of 4 nats in $T_0$ would be the difference between a
vaccine having $T_0=0$ and $E=0$ (makes no difference at all) and one
which has $T_0=-4$ and $E=0.96$, i.e. is 96\% effective; or
e.g. between one which has $T_0=-2$ and $E=0.76$, i.e. is 76\%
effective and one which has $T_0=+2$ and $E=-0.76$, i.e. one that
increases the incidence of covid by 76\%.

Equally if $j_0=0.05$ then it would be the difference between a
vaccine having $T_0=0$ and $E=0$ (vaccination makes no difference) and
one having $T_0=-4$ and $E=0.98$ (98\% effective), or between one
having $T_0=-2$ and $E=0.86$ (86\% effective) and one having $T_0=+2$
and $E=-4.6$ ($-460\%$ effective, or increases infections by 460\%).

An uncertainty of $4$ nats in $T_0$ would therefore render the
information obtained essentially useless. 

\subsection{Partial real-life data}
\label{realdataintro}

Instead of using synthetic data made from the model, we can also run
the model using real-life data if available. In order to check that
real-life data is not of a nature that results in a different
inference conclusion, we applied inference using the model to the data
from \cite{LopezBernalNEJM}, indeed specifically to those who were
either completely unvaccinated or had received two doses of the
vaccine, and to those who either tested negative or were infected with
the alpha variant of covid. Specifically this gave us the numbers
shown in table \ref{LB} covering a total of $N=127820$ patients seen
in hospital and tested.

\begin{table}[hpt]
\begin{center}
\begin{tabular}{c|c|c}
& \multicolumn{2}{c}{Test status}\\ \cline{2-3}
Vaccination status & Negative & Alpha positive\\\hline
Unvaccinated & $96371$ & $7313$\\\hline
Vaccinated & $23993$ & $143$\\\hline
\end{tabular}
\caption{Data on patients from table 2 of \cite{LopezBernalNEJM} who
  were tested either negative or positive for the alpha variant of
  covid, and who were either completely unvaccinated, or who had
  received two doses of any covid vaccine.
\label{LB}
}
\end{center}
\end{table}

Now, the uses of the model described above all assume that the
vaccination status of the entire population is known, even though the
covid-test status is only known for those who are hospitalised. In the
case of the real-life data from \cite{LopezBernalNEJM} we do not know
the vaccination status of the entire population. Therefore we
considered three possible assumptions about the unseen part of the
population, in all cases assumed to number $2N$, namely:

\begin{description}

\item[Assumption 1:] $19\%$ of the unseen population is vaccinated
  (the same fraction as in those seen in hospital);

\item[Assumption 2:] The entire unseen population is vaccinated; or

\item[Assumption 3:] Half of the unseen population are vaccinated or
  unvaccinated in the same proportion to those seen in hospital, while
  the vaccination status of the remainder is unknown.

\end{description}

\section{Results}

In all cases the reader who prefers to see percentage values of $E$
rather than plots of $T_0$ is referred to appendix \ref{Evalsappx}.

\subsection{The base case -- wide open priors}
\label{wideopensection}

\begin{figure}[hpt]
\begin{center}
\includegraphics[scale=0.5]{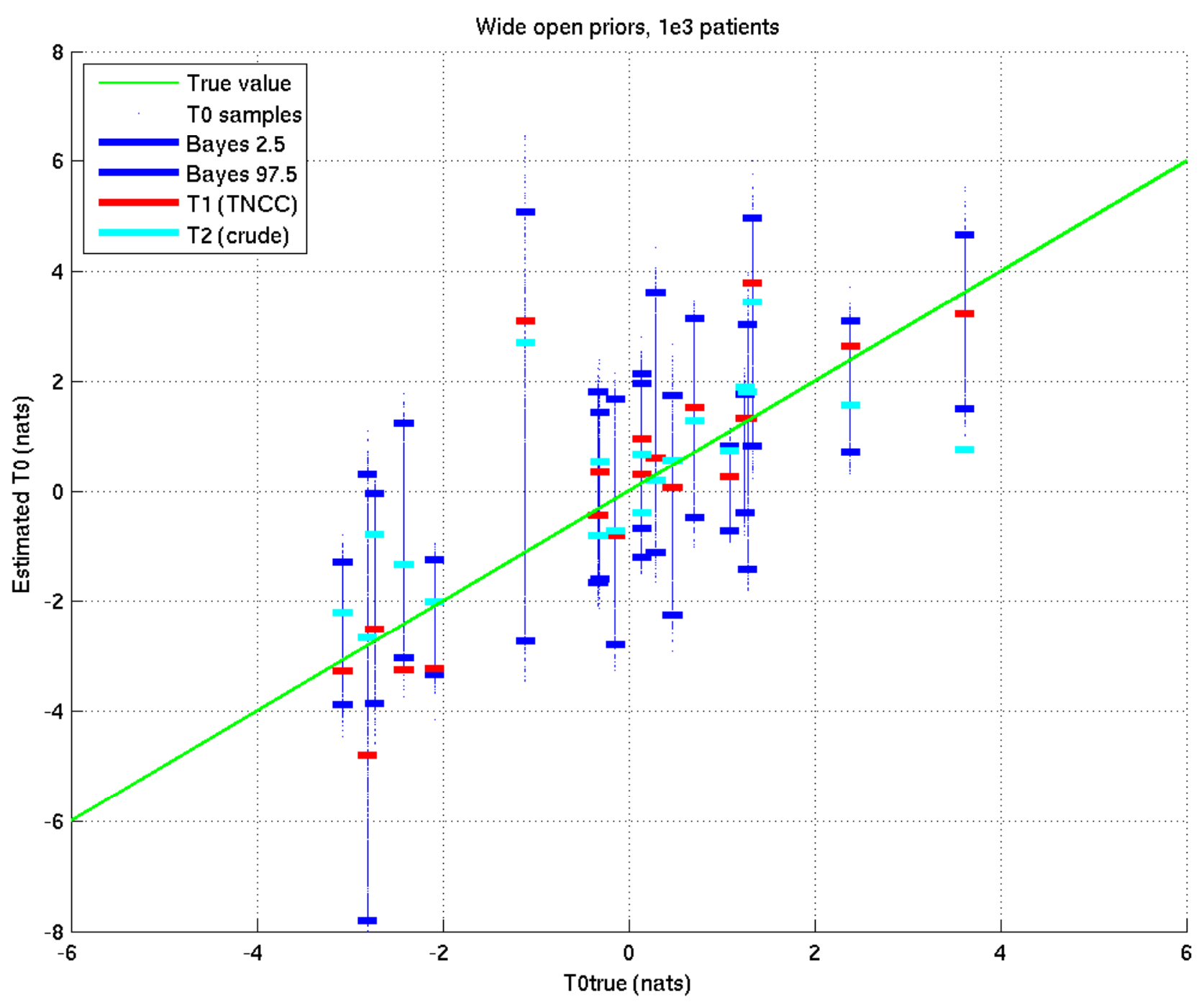}

\caption{The results of using the base-case, wide-open prior, with
  $N=1000$. If we do not know the values of the various probabilities
  comprising $p,r,j,q$ and believe them to be equally likely to take
  any values in $[0,1]$, then: about half the patients get
  hospitalised; of the 20 runs, 10 had $T_1$ more accurate than $T_2$;
  the standard deviation of the error was 1.29 nats for $T_1$ and 1.35
  nats for $T_2$; and the difference between the 2.5th and 97.5th
  centiles of the posterior for $T_0$ was 3.8 nats on average.
\label{wideopen}
}
\end{center}
\end{figure}

We start by looking at what happens in the base case when all the
priors are uniform on $[0,1]$ and we have $N=1000$ patients (the
default number except where stated otherwise). The results of the 20
random samples from the model are shown in figure
\ref{wideopen}. Where the red mark is nearer the green line than the
cyan mark is, TNCC is more accurate than the crude estimate, and vice
versa. The tiny blue dots are samples from the Bayesian posterior on
$T_0$, and where the two dark blue marks (the 2.5th and 97.5th
centiles) are close together the data determines $T_0$ more accurately
than when they are far apart.

We see from this that in this case there is essentially no difference
in quality between the TNCC ($T_1$) estimate and the crude estimate
($T_2$) -- both are fairly useless in determining vaccine efficacy,
with errors of around 1.3 nats standard deviation. But, even though on
average both of them are out by around 1.3 nats, the Bayesian
confidence intervals obtained have a width of 3.8 nats on average,
with some much wider than this -- so that the results are essentially
useless, even if the root-mean-square error is only 1.3 nats.

Those interested in checking the plausibility of the samples giving
the extreme values of $T_0$ are directed to appendix section
\ref{samples} below.

\subsection{The base case but with more patients}

\begin{figure}[hpt]
\begin{center}
\includegraphics[scale=0.5]{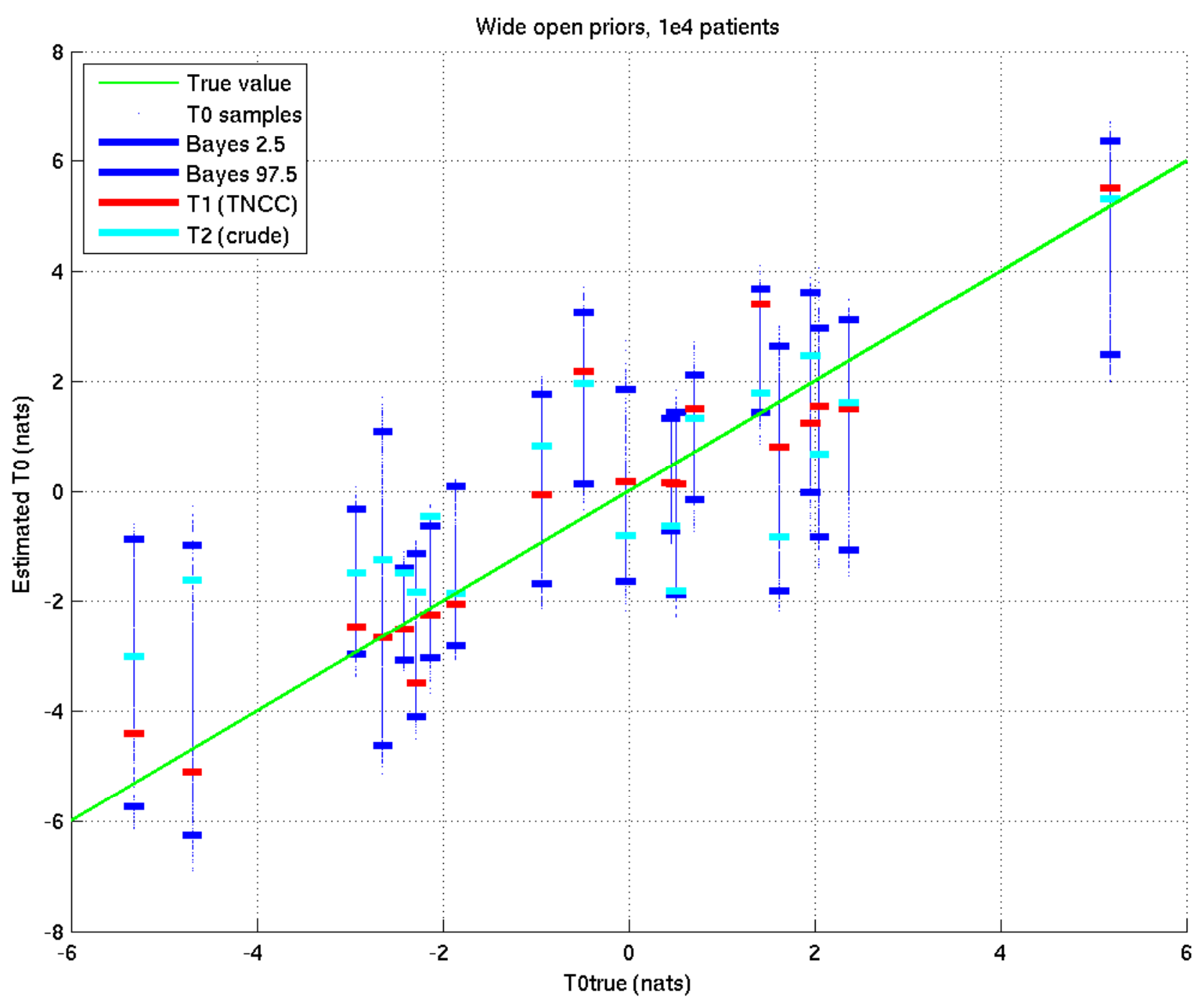}

\caption{The results of using the base-case, wide-open prior. If we do
  not know the values of the various probabilities comprising
  $p,r,j,q$ and believe them to be equally likely to take any values
  in $[0,1]$, but now have $N=10000$ patients, then: of the 20 runs,
  12 had $T_1$ more accurate than $T_2$; the standard deviation of the
  error was 0.94 nats for $T_1$ and 1.6 nats for $T_2$; and the
  difference between the 2.5th and 97.5th centiles of the posterior
  for $T_0$ was 3.4 nats on average.
\label{wideopenmore}
}
\end{center}
\end{figure}

We wonder next whether the huge uncertainty in the inference might be
because we don't have enough patients. We therefore increase the
number of patients to $N=10000$ patients, without changing the
wide-open priors. The results are then shown in figure
\ref{wideopenmore}. Again, about half the patients get hospitalised,
so we are not talking of very small numbers being observed.

Although the standard deviation of the TNCC error has fallen somewhat,
the width of the Bayesian confidence intervals are still far too large
to be of any practical use. Note also that ten thousand patients may
not seem like that many, but given that we are considering relatively
high hospitalisation rates, the data we are getting is probably
comparable to a catchment population of a million or so in real life
in the specific current covid situation.

\subsection{Prior 2 -- With almost everybody in the same subset of the
  population}

\begin{figure}[hpt]
\begin{center}
\includegraphics[scale=0.5]{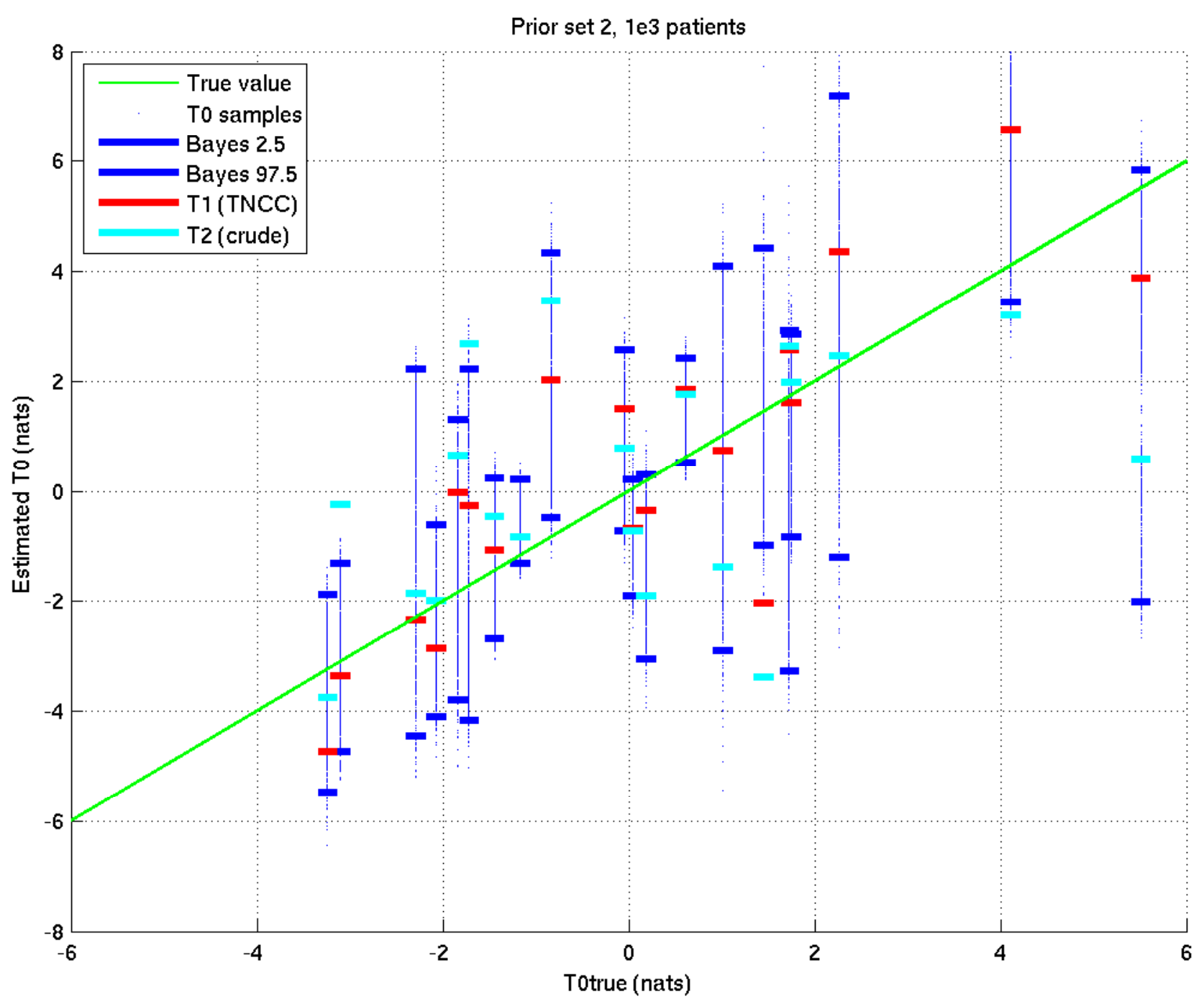}

\caption{The results of using Prior 2 with one subset containing
  $>99\%$ of the population. Then: of the 20 runs, 13 had $T_1$ more
  accurate than $T_2$; the standard deviation of the error was 1.5
  nats for $T_1$ and 2.4 nats for $T_2$; and the difference between
  the 2.5th and 97.5th centiles of the posterior for $T_0$ was 4.6
  nats on average.
\label{var2}
}
\end{center}
\end{figure}

In fact it turns out that we don't need two large subsets of the
population to cause these problems. Using Prior 2, we get figure
\ref{var2}.

We again see that it is pot-luck which of TNCC or crude estimate does
better, and that the Bayesian confidence intervals are still wide open
on average.

\subsection{Prior 1 -- Tight priors which make TNCC consistently better than the
  crude estimate}

\begin{figure}[hpt]
\begin{center}
\includegraphics[scale=0.5]{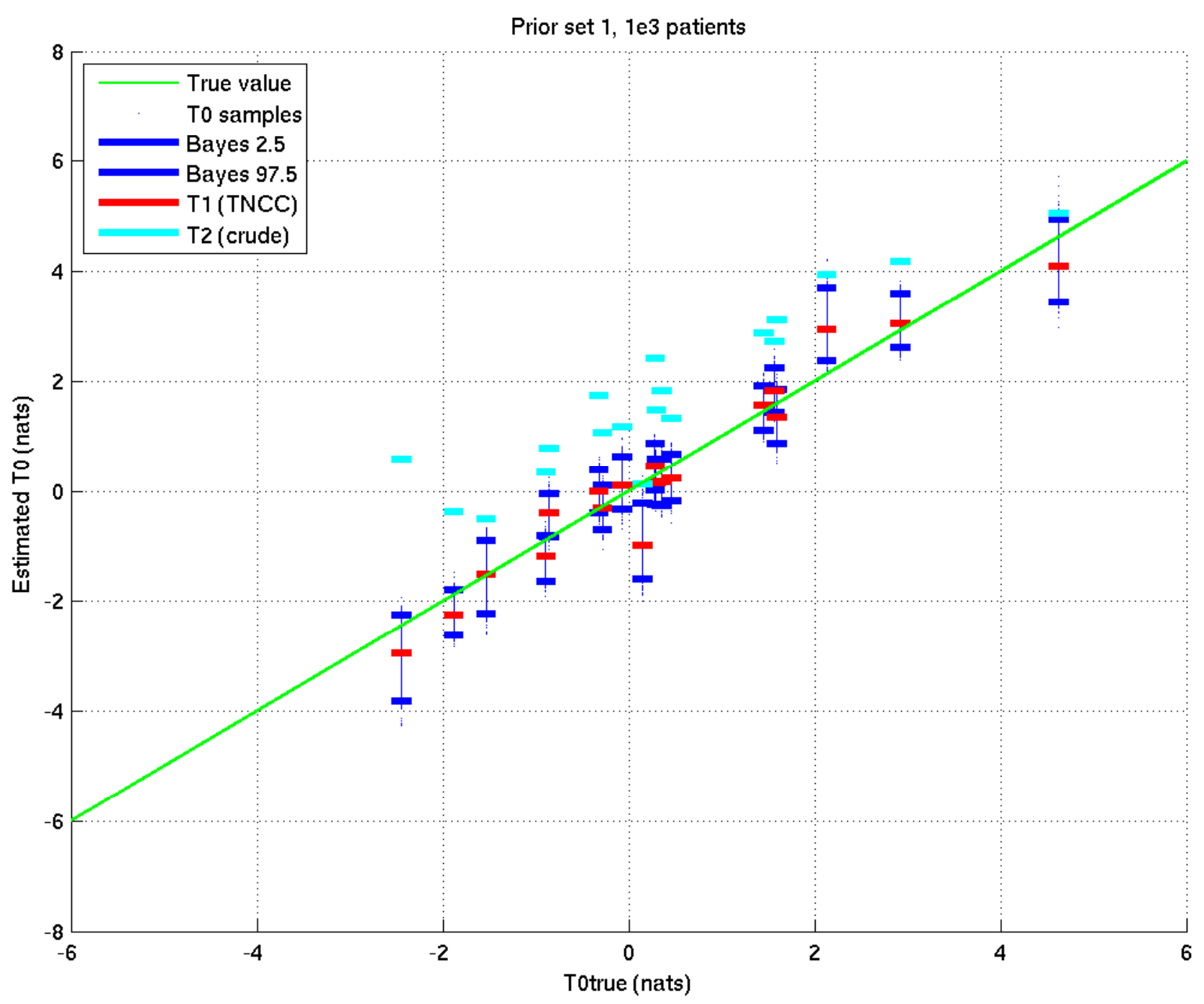}

\caption{The results of using Prior 1, using $N=1000$ patients. Then:
  about 697 patients get hospitalised on average; of the 20 runs, 18
  had $T_1$ more accurate than $T_2$, with all the $T_2$ estimates
  being too high; the standard deviation of the error was 0.44 nats
  for $T_1$ and 1.96 nats for $T_2$; and the difference between the
  2.5th and 97.5th centiles of the posterior for $T_0$ was 1.1 nats on
  average.
\label{var1}
}
\end{center}
\end{figure}

However, there are other settings of the priors which make the TNCC
method consistently better than the crude estimate. Using Prior 1 we
get figure \ref{var1}.

So in this case we are saying that we know the values of the various
probabilities affecting whether people are vaccinated or not and
whether or not they are hospitalised under the various circumstances,
and that we know them fairly accurately. Then -- having made this very
strong assumption -- TNCC is much better than the crude estimate, and
moreover the Bayesian confidence interval is much narrower than with
wide open priors. Indeed, the crude estimate is now often even some
distance outside the Bayesian confidence interval.

\subsection{Prior 1 and more patients -- Tight priors which make
  TNCC even better}

\begin{figure}[hpt]
\begin{center}
\includegraphics[scale=0.5]{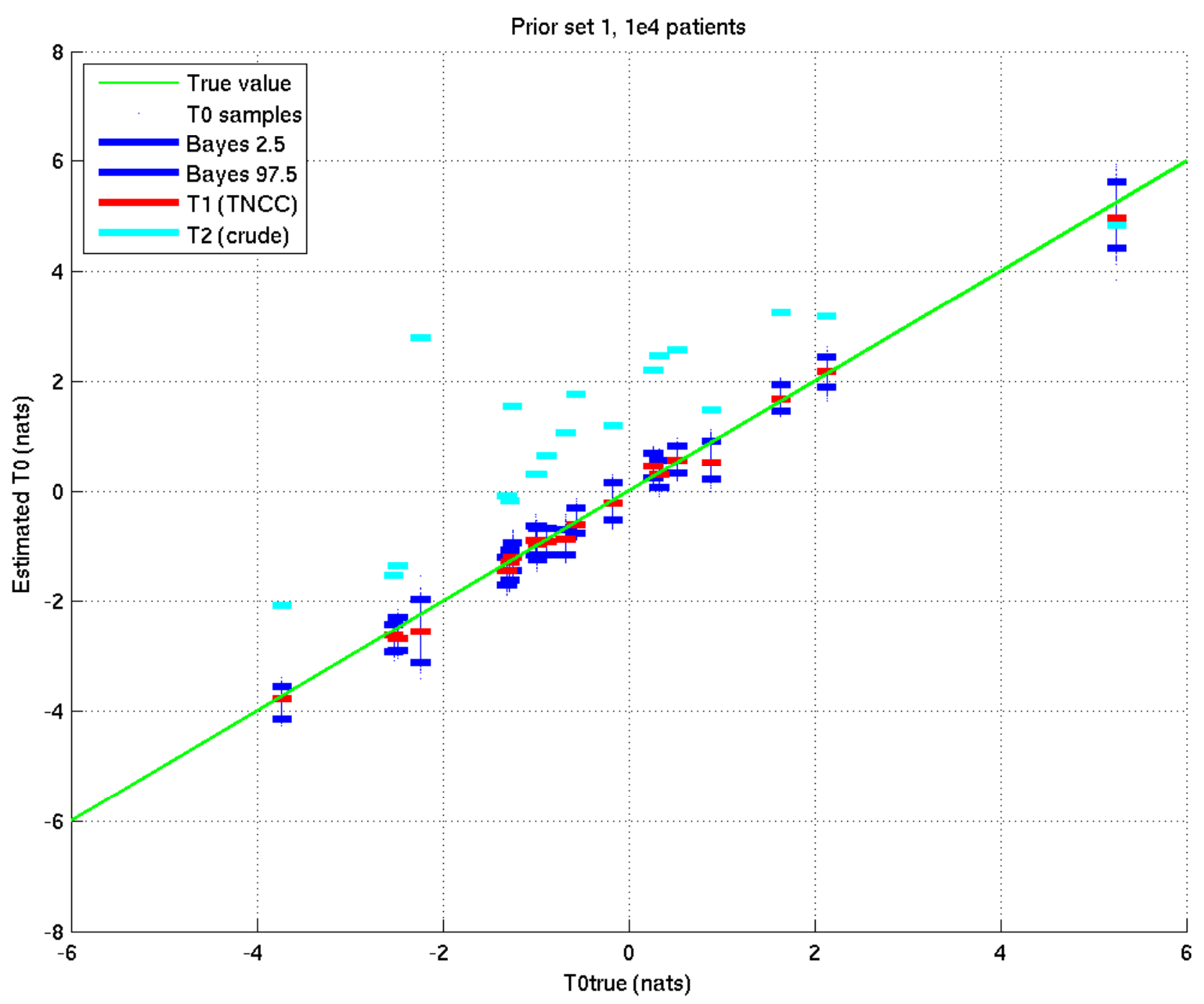}

\caption{The results of using Prior 1 but now with $N=10000$
  patients. Then: of the 20 runs, all had $T_1$ more accurate than
  $T_2$, with all the $T_2$ estimates being too high; the standard
  deviation of the error was 0.16 nats for $T_1$ and 1.9 nats for
  $T_2$; and the difference between the 2.5th and 97.5th centiles of
  the posterior for $T_0$ was 0.6 nats on average.
\label{var1more}
}
\end{center}
\end{figure}

In this case we can do even better by looking at a larger number of
patients. If we set $N=10000$ and again use Prior 1, we get figure
\ref{var1more}.

So in this case we are again saying that we know the values of the
various probabilities affecting whether people are vaccinated or not
and whether or not they are hospitalised under the various
circumstances, and that we know them fairly accurately, and that we
have a lot more patients (and a lot more being hospitalised). Then --
having made this very strong assumption -- TNCC is even better than
it was with fewer patients, and moreover the Bayesian confidence interval is
even narrower than with fewer patients.

\subsection{Prior 3 -- Tight priors which make TNCC worse than the
  crude estimate}

\begin{figure}[hpt]
\begin{center}
\includegraphics[scale=0.5]{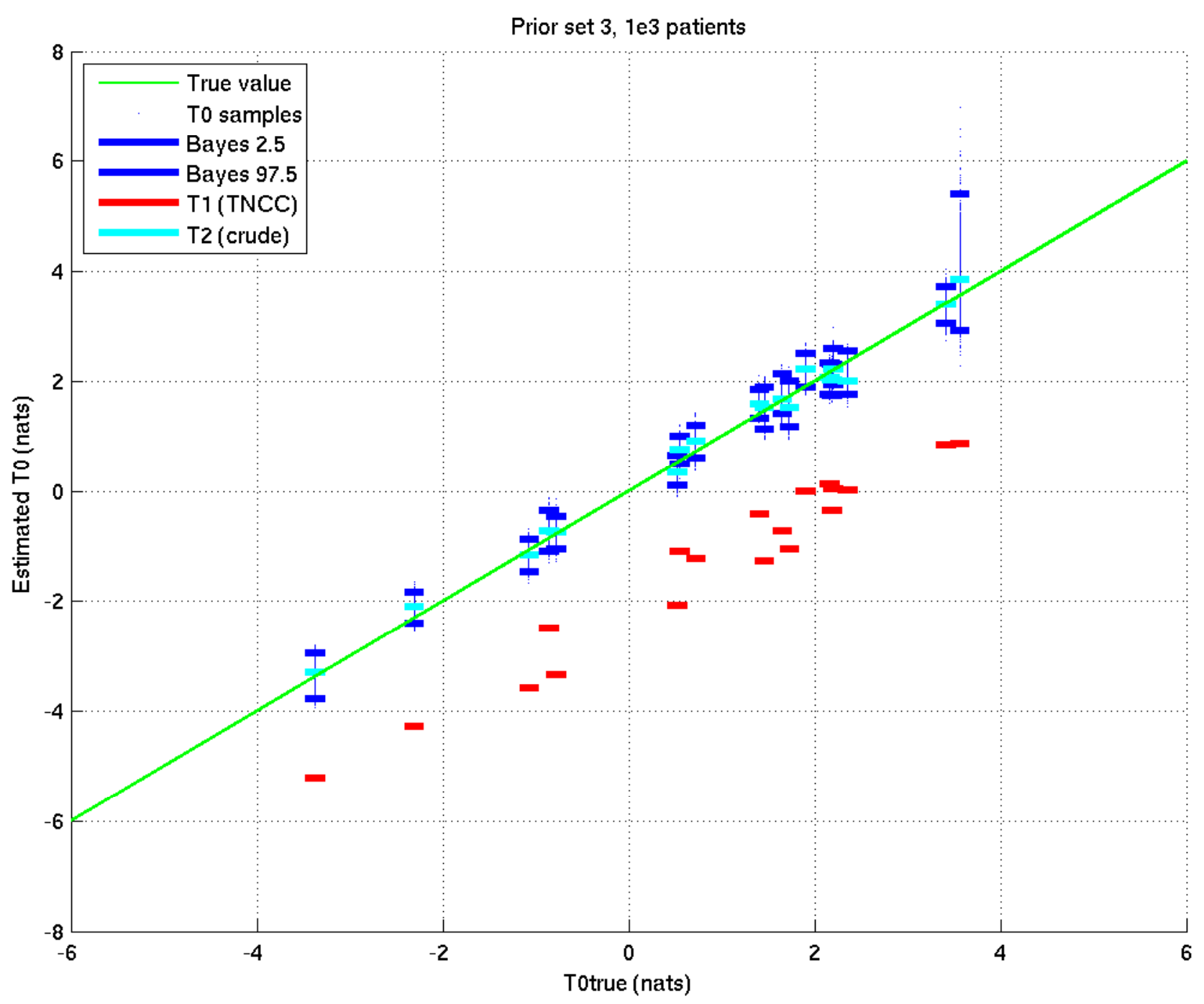}

\caption{The results of using Prior 3 (with $N=1000$ patients). Then:
  on average around 790 patients get hospitalised; of the 20 runs,
  \textit{none} had $T_1$ more accurate than $T_2$, with all the $T_1$
  estimates being too low (i.e. suggesting vaccines are more effective
  than they are); the standard deviation of the error was 2.4 nats for
  $T_1$ and 0.71 nats for $T_2$; and the difference between the 2.5th
  and 97.5th centiles of the posterior for $T_0$ was 0.85 nats on
  average.
\label{var3}
}
\end{center}
\end{figure}

However, it also entirely possible to have tight priors that make TNCC
much \textit{worse} than the crude estimate. If we use Prior 3, we
get figure \ref{var3}.

In this case we are again saying that we know the values of the
various probabilities fairly accurately, but that they have different
values than with the priors previously tried. In this case the crude
estimate is far \textit{better} than TNCC; indeed in every case TNCC
is now way outside the Bayesian confidence interval.

\subsection{With prior 0 and partial real-life data}

Under the three possible assumptions on the unseen part of the
population described in section \ref{realdataintro}, we obtained the
results shown in table \ref{LBresults}. 

\begin{table}[hpt]
\begin{center}
\begin{tabular}{c|c|c|c|c|c}
Assumption & $T_1$ & $T_2$ & $T_0$ $2.5\%$ile & $T_0$ $97.5\%$ile & $N_{\text{hosp}}$\\\hline
1 & -2.54 & -2.49 & -3.53 & +3.11 & 127820\\\hline
2 & -2.54 & -5.00 & -4.79 & +3.46 & 127820\\\hline
3 & -2.54 & -2.49 & -4.81 & +0.41 & 127820\\\hline
\end{tabular}
\caption{Results of inference using real-life data as in table
  \ref{LB} and the three assumptions listed in section
  \ref{realdataintro}. 
\label{LBresults}
}
\end{center}
\end{table}

In all three cases, the width of the posterior distribution between
the $2.5$th and $97.5$th centiles is greater than $5$ nats, confirming
that in this case also one cannot tell to any useful extent how
effective the vaccine actually is. 

\subsection{Summary of results}

Summarising the results, we see that the data obtained, with wide open
priors, does not contain enough information to enable us to determine
vaccine effectiveness to any useful extent, whether we use simulated
or real-life data. Moreover there is very little to choose between the
TNCC and crude estimates, which are each better than the other about
half of the time, and increasing the number of patients does little to
narrow the Bayesian confidence intervals.

On the other hand if we have solid information on the various prior
probabilities comprising $p,r,j,q$, then, depending on those values,
it may be that the TNCC method becomes far better than the crude
method (and that the data then easily contains enough information
about vaccine effectiveness to be useful) and improves still further
with more patients; or on the other hand it may be such that the crude
method becomes far better than the TNCC method (and again the data
contains useful amounts of information).

Note, however, that we are \textit{not} saying that where the crude
method is right, TNCC is always too optimistic, or that where TNCC is
right, the crude method is always too pessimistic. Indeed, the setup
is symmetric in the $l$ dimension, so if one swaps $l=0$ and $l=1$ in
the relevant variant's prior the opposite error will become evident.

So in summary, unless we have priors that are both tight, and
favourable to TNCC, TNCC is not reliably an improvement on the crude
method -- even if the population is essentially homogeneous. In the
usual setting we do not have tight priors on these probabilities, let
alone ones we know to be favourable to TNCC.

\section{Discussion}

We have shown that in the absence of informative priors, even in a
homogenous population, data collected by monitoring rates of
vaccinated and unvaccinated, covid-infected and uninfected, in only
the hospitalised (or only the tested) population, along with whole
population vaccination status, does not contain sufficient information
to determine with any useful accuracy the extent to which vaccination
prevents infection. This is in no way a covid-specific finding, and
indeed it has been theoretically foreseen in \cite{Sullivan} and
\cite{Westreich} years before the covid pandemic. Nonetheless, the
same conclusions apply to real-life data collected in respect of the
covid pandemic.

Neither the TNCC method nor the crude method is clearly better than
the other; in some circumstances one is more accurate, and in others
the other is more accurate, and we cannot tell which without knowing
details of the population's behaviour that are very hard to measure --
indeed in all circumstances where this behaviour remains unknown,
neither method is useful to any appreciable extent.

Sadly we notice that there is a tendency to assume that the method
giving the results we prefer is correct, for example in \cite{Week48}
and in daily BBC television news in the UK preferring the TNCC result
indicating that vaccination is highly effective, while GB news
(another UK TV channel) prefers the result of the crude method
\cite{GBnews}, indicating that vaccination actually increases the
incidence of covid.

However these facts do not reduce the importance of ongoing monitoring
of vaccine effectiveness and side effects. New variants of
infectious agents arise frequently and may affect vaccine
effectiveness; manufacturing changes may affect the incidence of
side-effects, and even if not, when larger numbers are vaccinated more
side-effects are likely to come to light. Even where individual
side-effects are not statistically significant, the overall rate of
serious side-effects may become both statistically significant and
significant to the healthcare of the population.

\section{What should we do about it ?}

We can suggest two ways forward that do achieve reliable monitoring of
vaccine effectiveness, both of which involve more work than just
sitting back and waiting for data to flow to us:

\begin{enumerate}

\item \label{randomtest} Organise the testing randomly, independent of
  vaccination status or perception of being ill. By continually
  picking random members of the community to be tested, and then
  \textit{exhaustively} ensuring that you get all the results back
  (whether they are at home / on holiday / in hospital) -- which is a
  lot of work -- you can ensure that being hospitalised (i.e. being
  tested) is conditionally independent of being vaccinated given your
  infection status ($h\perp v|l$), which as we show in appendix
  section \ref{condindrels} ensures that you can easily deduce the
  vaccine effectiveness measure $T_0$.

\item \label{doRCTs} Do ongoing randomised controlled trials (RCTs) on
  a randomly chosen subset of the population. This obviously requires
  patient consent, and has the drawback (for those who believe a
  particular vaccine works well) that not everybody gets vaccinated
  that wants to be. Follow-up is again critical, which makes it hard
  work. It has, however, the big advantage that it also allows ongoing
  monitoring for side-effects which haven't been detected during the
  pre-deployment clinical trials.

\end{enumerate}

Therefore it is our opinion that when vaccines are being given to
large numbers of people it is important that ongoing double-blind,
randomised, placebo-controlled clinical trials continue, as properly
designed RCTs can avoid all the uncertainty to do with not knowing how
patients decide to come forward for testing\cite{RCTreview}. This can
be done, for example, by enrolling (subject to informed consent) a
randomly selected subset of the population for randomisation to
vaccine or placebo and following them up long term. Such RCTs should
not exclude any part of the population offered vaccination and able to
give informed consent, lest the results fail to correctly represent
actual practice. Further, they should be conducted by investigators
unconnected with the commercial interests of the vaccine manufacturers
or the political interests of the government and its advisers, and
without any restriction on publication of the results.

In the specific case of the covid pandemic and the vaccines available
as of mid 2022, it is notable that the trials undertaken
(\cite{Pfizer1, Pfizer2, Pfizer3, AZvaccine}) have been of relatively
low numbers and relatively short duration. Moreover the potential for
ongoing collection of randomised data from at least \cite{Pfizer2} has
been reduced by offering all the placebo-recipients vaccine when
emergency authorisation of the vaccine was granted.  There have also
been a number of reports casting doubt on the validity of the results
of these trials (e.g. \cite{Doshi}, \cite{Ventavia}). The duration of
any protection given by the vaccines, and their activity against
variants, is also in doubt. The use of the TNCC method in ongoing
monitoring is being preferred over the crude method without any
justifying work showing that the underlying probabilities can be
assumed to be sufficiently favourable to TNCC. We therefore think that
it is especially important in this case that ongoing RCTs are
undertaken.

\appendix

\section{Appendix - Definitions}
\label{defs}

\subsection{Notation}

We define the following variables:
\begin{itemize}

\item $v$ is the state of vaccination of a patient, 1 if vaccinated
  and 0 if not. If we want to refer to the vaccination state of
  patient $n$ then we will refer to $v_n$, and similarly for the
  following variables.

\item $l$ is whether or not a patient is infected with covid, 1 if so,
  and 0 if not.

\item $h$ is whether or not a patient is hospitalised (i.e. tested), 1
  if so, and 0 if not.

\item $s$ indicates which of two subsets 0 and 1 of the population a patient
  belongs to, which differ only in the not-directly-observable nature
  of their healthcare-seeking behaviour.

\end{itemize}

We assume throughout that, within each of the two $s$ subsets of the
population, the population is entirely homogeneous. This is of course
a gross simplification in comparison with reality; but if, even with
this simplification, the inference is uncertain, then it is reasonable
to assume that it is uncertain in reality also. (Indeed our results
show that inference is equally uncertain even if the population is
almost entirely homogeneous rather than in two differing subsets.)
Similarly we assume that being tested for covid is synonymous with
being hospitalised. 

To save symbols, we will use $V$ to denote the state that $v=1$, and
use e.g. $P(V)$ to mean $P(V=1)$ and $P(v)$ to mean $P(v=0)$. Thus for
example $P(vH|l)$ means the probability that a patient is hospitalised
and hasn't been vaccinated given that they do not have covid.

The symbol $\#$ will denote one plus the number of individuals in a
particular subset. For example, $\#lVH$ denotes one plus the number of
individuals who have been vaccinated and are in hospital but do not have
covid. (The reason for the ``one plus'' is so that
e.g. $\frac{\#LV}{\#lV + \#LV}$ gives the posterior expectation of
$P(L|V)$ given flat priors.)

\subsection{Definitions of measures of vaccine effectiveness}

We use the following further notation in relation to measures of
vaccine effectiveness (all logarithms being natural, i.e. to base
$e$):
\begin{itemize}

\item $E$, also denoted $E_0$, is the usual definition of vaccine
  effectiveness, given by $$E = 1 - \frac{P(L|V)}{P(L|v)}.$$

\item $T_0$ is a slightly different measure of vaccine effectiveness,
  defined by $$T_0 = \log\left(\frac{P(L|V)}{1 - P(L|V)} /
  \frac{P(L|v)}{1 - P(L|v)}\right),$$ the difference between the log
  odds of having covid if you are vaccinated and that log odds if you
  are unvaccinated. We note that also $$T_0 =
  \log\left(\frac{P(L|V)}{P(l|V)} / \frac{P(L|v)}{P(l|v)}\right).$$
  Thus a very negative value of $T_0$ goes with vaccination being
  effective, but $T_0$ will be the same if vaccination reduces the
  probability of getting covid from $0.2$ to $0.1$ as if it increases
  the probability of \textit{not} getting covid from $0.1$ to $0.2$,
  unlike $E$, which changes much less in the latter situation. The
  reason for being interested in $T_0$ is that it intuitively
  corresponds closely with what is measured by TNCC. It is also easier
  to appreciate in plots when probabilities are close to zero or one.

\item $T_1$ is what is measured by TNCC. If this were a
  non-homogeneous population in which various other covariates were
  also measured, it would be calculated by logistic regression, but in
  this population in which all directly observable characteristics are
  the same for everybody, simple division gives us what we need. We
  thus define 
  \begin{IEEEeqnarray*}{lCr}
  T_1 &=& \log\left(\frac{\#LHV}{\#LHv} / \frac{\#lHV}{\#lHv}\right)\\
  &=& \log\left(\frac{\#LHV}{\#lHV} / \frac{\#LHv}{\#lHv}\right).
  \end{IEEEeqnarray*}
  Comparison of this with the definition of $T_0$ makes the intuitive
  correspondence clear.

\item $T_2$ is the ``obvious'' crude estimate of $T_0$, given by 
$$T_2 = \log\left(\frac{\frac{\#LHV}{\#V} / \left(1 - \frac{\#LHV}{\#V}\right)}
                       {\frac{\#LHv}{\#v} / \left(1 - \frac{\#LHv}{\#v}\right)}\right).$$
To get an intuitive feel for this, note that if we were going for
estimating $1-E$ rather than $T_0$, we would instead be looking at
$$Q=\frac{\frac{\#LHV}{\#V}}{\frac{\#LHv}{\#v}},$$ the ratio of the
infection rate in the vaccinated population to that in the
unvaccinated. Thus $T_2$ is to $T_0$ for odds what $Q$ is to $1-E$ for
probabilities.

\item $E_1$ and $E_2$ respectively correspond to $T_1$ and $T_2$ as
  $E_0$ does to $T_0$, using the equation below and the estimates of
  $P(L|v)$ given at the start of appendix \ref{Evalsappx}.

\end{itemize}

We note in passing that if we know the baseline probability $P(L|v)$
of an unvaccinated individual getting covid then we can determine
either of $E_0$ and $T_0$ from the other, using the equation $$E_0 = 1 -
\frac{e^{T_0}}{1 + (e^{T_0} - 1)P(L|v)}.$$

\subsection{Conditional independence relationships}

\label{condindrels}

We write $a\perp b | c$ to mean ``$a$ is conditionally independent of
$b$ given $c$'', and note the following points:

\begin{enumerate}
\item Considering the TNCC estimate $T_1$:
\begin{IEEEeqnarray*}{rCl}
e^{T_1} & \approx & \frac{P(LHV)}{P(lHV)} / \frac{P(LHv)}{P(lHv)}\\
& = & \frac{P(H|LV)P(L|V)P(V)}{P(H|lV)P(l|V)P(V)} /
\frac{P(H|Lv)P(L|v)P(v)}{P(H|lv)P(l|v)P(v)}\\
& = & \frac{P(H|LV)P(H|lv)}{P(H|Lv)P(H|lV)}
\left(\frac{P(L|V)}{P(l|V)} / \frac{P(L|v)}{P(l|v)}\right)\\
& = & \frac{P(H|LV)P(H|lv)}{P(H|Lv)P(H|lV)}e^{T_0}
\end{IEEEeqnarray*}
so that $T_0=T_1$ if and only
if $$\frac{P(H|LV)P(H|lv)}{P(H|Lv)P(H|lV)} = 1,$$ which will be the
case in particular if either $h\perp l|v$ or $h\perp v|l$, neither of which
conditional independencies is remotely likely.

\item Similarly, let us assume that the vast majority of the
  vaccinated who are not both hospitalised and infected are in fact
  neither hospitalised nor infected, and similarly for the
  non-vaccinated, i.e. that $$P(lh|V)\gg P(Lh|V) +
  P(lH|V)$$ $$P(lh|v)\gg P(Lh|v) + P(lH|v),$$ which is of course true
  for most infectious diseases (and was even during the covid
  pandemic for covid). Then we find that $T_0=T_2$ if and only
  if $$\frac{P(H|LV)P(h|lv)}{P(h|lV)P(H|Lv)}=1,$$ which again will be
  the case in particular if $h\perp v|l$, which is one of the same
  conditional independencies as in the previous point. 

\item Now, the conditional independency that made $T_1\approx T_0$
  that doesn't achieve the same for $T_2 \approx T_0$ is $h\perp l|v$,
  i.e. that, both in the vaccinated and unvaccinated populations,
  getting hospitalised (i.e. tested) is independent of whether or not
  you are infected. It would be astonishing for this to be true.

  So there is really nothing to choose between $T_1$ and $T_2$ in this
  regard, unless one has reason to believe that, in the absence of any
  conditional independencies, one of these ratios is more likely to be
  unity than the other.

\item Since in real life we observe different values of $T_1$ and
  $T_2$, it follows that $h\not\perp v|l$.

\item If it were the case that $h\perp (l,v)$ then it would follow
  that $h\perp v | l$, and hence that $T_0=T_1=T_2$. Since we know
  that $h\not\perp v|l$, we also know that -- at least spontaneously
  -- we do not have $h\perp (l,v)$. However, it would be possible to
  collect a different dataset in which we did have $h\perp (l,v)$,
  remembering that the key point about the variable $h$ here is that
  it measures whether the patient is \textit{tested} for infection. 

  If, instead of just waiting for data to flow to us spontaneously, we
  deliberately send out tests to random members of the population
  (irrespective of vaccination status) and ensure that we get
  \text{all} the results back (which is really hard work !), then we
  would have a dataset in which $h\perp (l,v)$, which would provide
  one route forward to getting reliable monitoring data. It would then
  make no difference whether we analysed it believing it to be TNCC
  data or crude data.

\end{enumerate}

\section{Appendix - Model used and simulation and inference methods}
\label{model}

\subsection{Additional variable definitions}

We first make the following definitions of variables that are not
directly observable:
\begin{itemize}

\item $p$ is the probability that an individual belongs to the subset
  with $s=1$, i.e. $p = P(S)$.

\item $\alpha = (\alpha_0,\alpha_1)$ are parameters for a Beta prior
  distribution on $p$. Specifically $$P(p|\alpha) =
  \frac{\Gamma(\alpha_0+\alpha_1)}{\Gamma(\alpha_0)\Gamma(\alpha_1)}(1-p)^{\alpha_0-1}p^{\alpha_1-1},$$
  where $\Gamma$ denotes the Gamma function\footnote{For positive
    integers $m$, $\Gamma(m) = (m-1)!$; otherwise $\Gamma(x)$ is the
    natural mathematical extension of the function giving the
    factorial of $x-1$ to non-integers.}. We note that in
  particular if $\alpha=(1,1)$ then the prior on $p$ is uniform on
  $[0,1]$, and that if $\alpha_0$ and $\alpha_1$ are large then the
  prior is narrow with a mean of $\frac{\alpha_1}{\alpha_0+\alpha_1}$.

\item $r=(r_0,r_1)$ are the probabilities that an individual with $s=0,1$
  respectively is vaccinated. Thus $r_0=r_s=P(V|s)$.

\item $\beta = ((\beta_{0,0},\beta_{0,1}),\,(\beta_{1,0},\beta_{1,1}))
  = ((\beta_{s,0},\beta_{s,1}))_{s\in\{0,1\}}$ are the parameters for
  the Beta priors on $r_0,r_1$ respectively.

\item $j=(j_0,j_1)$ are the probabilities that an individual who isn't
  or is vaccinated respectively gets covid. Thus $j_1=j_V=P(L|V)$.

\item $\gamma = ((\gamma_{0,0},\gamma_{0,1}),\,(\gamma_{1,0},\gamma_{1,1}))
  = ((\gamma_{v,0},\gamma_{v,1}))_{v\in\{0,1\}}$ are the parameters for
  the Beta priors on $j_0,j_1$ respectively.

\item $q = (q_{s,l,v})_{s,l,v\in \{0,1\}}$ are the probabilities
  that an individual with the given values of $s$, $l$, and $v$ is
  hospitalised. For example $q_{0,0,1}=q_{slV}=P(H|slV)$.

\item $\delta =
  ((\delta_{s,l,v,0},\delta_{s,l,v,1}))_{s,l,v\in\{0,1\}}$ are the
  parameters for the Beta priors on the various components of
  $q$. Thus the parameters for the prior on $q_{Slv}$ are
  $(\delta_{1,0,0,0},\delta_{1,0,0,1})$.

\end{itemize}

\subsection{Model construction}

We simulate the population events over some fixed time period by
generating random samples from the following model. For each sample we
work out the true value of $T_0$, the TNCC and crude estimates $T_1$
and $T_2$, and take many (1000 or 10000) samples from the posterior
distribution on $T_0$ to establish how accurately it is (or is not)
possible to infer $T_0$ from the available data.

Intuitively we generate a random sample from our model by 
\begin{itemize}
\item first drawing $p$ from its Beta prior, 

\item then drawing $s_n$ for each patient $n$ from the discrete
  distribution on $\{0,1\}$ with probability $p$ of being 1, thus
  assigning each patient to one of the two subsets differing in
  healthcare-seeking behaviour;

\item then drawing $r$ (each component independently) from its Beta
  priors,

\item then drawing $v_n$ for each patient $n$ from the discrete
  distribution on $\{0,1\}$ with probability $r_{s_n}$ of being 1,
  thus determining whether each patient gets vaccinated or not,

\item then drawing $j$ (each component independently) from its Beta
  priors, 

\item then drawing $l_n$ for each patient $n$ from the discrete
  distribution on $\{0,1\}$ with probability $j_{v_n}$ of being 1,
  thus determining whether each patient gets covid or not,

\item then drawing $q$ (each component independently) from its Beta
  priors, 

\item then drawing $h_n$ for each patient $n$ from the discrete
  distribution on $\{0,1\}$ with probability $q_{s_n,l_n,v_n}$ of
  being 1, thus determining whether each patient is hospitalised,
  dependent on their subset, covid-status, and vaccination-status.

\item Finally we observe each patient's vaccination status, and, for
  those who are hospitalised only, whether or not they have covid.

\end{itemize}

Mathematically we define a standard Bayesian generative model for $N$
patients as follows: for $n=1,...,N$:
$$P(p|\alpha) =
\frac{\Gamma(\alpha_0+\alpha_1)}{\Gamma(\alpha_0)\Gamma(\alpha_1)}(1-p)^{\alpha_0-1}p^{\alpha_1-1};$$
$$P(S_n|p) = p;$$
$$P(r_s|\beta_{s,0},\beta_{s,1}) =
\frac{\Gamma(\beta_{s,0}+\beta_{s,1})}{\Gamma(\beta_{s,0})\Gamma(\beta_{s,1})}(1-r_s)^{\beta_{s,0}-1}r_s^{\beta_{s,1}-1};$$ 
$$P(V_n|s_n,r_{s_n}) = r_{s_n};$$
$$P(j_v|\gamma_{v,0},\gamma_{v,1}) =
\frac{\Gamma(\gamma_{v,0}+\gamma_{v,1})}{\Gamma(\gamma_{v,0})\Gamma(\gamma_{v,1})}(1-j_v)^{\gamma_{v,0}-1}j_v^{\gamma_{v,1}-1};$$ 
$$P(L_n|v_n,j_{v_n}) = j_{v_n};$$
$$P(q_{s,l,v}|\delta_{s,l,v,0},\delta_{s,l,v,1}) =
\frac{\Gamma(\delta_{s,l,v,0}+\delta_{s,l,v,1})}{\Gamma(\delta_{s,l,v,0})\Gamma(\delta_{s,l,v,1})}(1-q_{s,l,v})^{\delta_{s,l,v,0}-1}q_{s,l,v}^{\delta_{s,l,v,1}-1};$$ 
$$P(H_n|s_n,l_n,v_n,q_{s_n,l_n,v_n}) = q_{s_n,l_n,v_n};$$
where here the capitalised variables specifically denote the relevant
variable being 1, while the lower case range over 0 and 1 consistently
in each equation.

For the simulations we then consider the values of all $v_n$ to be
observed, along also with the values of $l_n$ for those $n$ for which
$h_n=1$. For some of the runs using partial real-world data from
\cite{LopezBernalNEJM} we consider some of the values of $v_n$ for
which $h_n=0$ to be unobserved, as described in the main text.

\subsection{Priors considered}
\label{priordescriptions}

Table \ref{priordefs} gives the details of the various priors
used. All are from the Beta family, and where a variable has several
components, each is considered to be independently distributed; except
where indicated otherwise all components have the same prior. Each
prior is either uniform on $[0,1]$ (i.e. has both parameters equal to
1) or has parameters summing to 200 such that
$\frac{\theta_1}{\theta_0+\theta_1}$ is equal to the value given which
is then the mean of the distribution, where $\theta$ denotes either
$\alpha,\beta_s,\gamma_v,$ or $\delta_{s,l,v}$.

\begin{table}[hpt]
\begin{center}
\begin{tabular}{l|c|c|c|c}
\textbf{Prior Name} & $p$ & $r_s$ & $j_v$ & $q_{s,l,v}$\\ \hline
Wide open & uniform & uniform & uniform & uniform \\ \hline
Prior 1 & 0.5 & $\begin{matrix}0.005 & (s = 0)\\ 0.995 & (s = 1)\end{matrix}$ & uniform & $\begin{matrix}0.4 &
  (s=0)\\ 0.995 & (s=1)\end{matrix}$\\ \hline
Prior 2 & 0.995 & uniform & uniform & uniform\\ \hline
Prior 3 & 0.995 & 0.5 & uniform & $\begin{matrix}0.995 & (l=1\text{ or }v=1)\\0.1 & (l=v=0)\end{matrix}$\\ \hline
\end{tabular}
\caption{Definitions of the various Beta priors used. Where the prior
  is not uniform, the value given is the mean of the relevant Beta
  distribution, the sum of whose parameters is 200.
\label{priordefs}
}
\end{center}
\end{table}

\subsection{Model evaluation and inference}
\label{MCMC}

In order to then infer the posterior distribution on $T_0$, we carry
out Markov chain Monte Carlo Gibbs sampling, visiting the different
unobserved variables (including those $l_n$ for which $h_n=0$) in
palindromic order, maintaining detailed balance as we go. The details
of how to do this are well described in \cite{rnealMCMC}, or see
\cite{Dagpunar} or appendix 2 of \cite{WhitworthLifetime}. To check
convergence we compare the time-course and distribution of the various
variables when started from values randomly generated by the model
with the time-course and distribution when started from the true
values that gave the observed data; in all cases these were
sufficiently similar not to change our conclusions. With $N=1000$
patients, 1000 samples were sufficient for convergence, while with
$N=10000$ patients approximately 10000 samples were needed. For the
real-world data with $127820$ patients whose test results were known
we used $200000$ samples.

Having obtained each random sample from the model, we evaluate the
corresponding value of $T_0$ by $$T_0 = \log\left( \frac{j_1}{1-j_1} /
\frac{j_0}{1-j_0}\right),$$ and similarly evaluate $T_{0,\text{true}}$
using the true values of $j$. The observed values of $T_1$ and $T_2$
can be calculated from their definitions by simply counting up the
number of patients with each set of attributes. In the case of runs
using real-world data from \cite{LopezBernalNEJM} in which some $v_n$
are unobserved, we use the mean posterior value of the number of
unvaccinated members of the population for calculating $T_2$. The
values of the centiles of $T_0$ are calculated as the centiles of the
set of samples of $T_0$ after excluding the first $10\%$ of samples as
run-in, and similarly the centiles of $E_0$ and the posterior means of
$T_0$ and $E_0$.

\section{Appendix - Details of posterior samples giving high and low values of
  $T_0$}
\label{samples}

\begin{figure}[hpt]
\begin{center}
\includegraphics[scale=0.5]{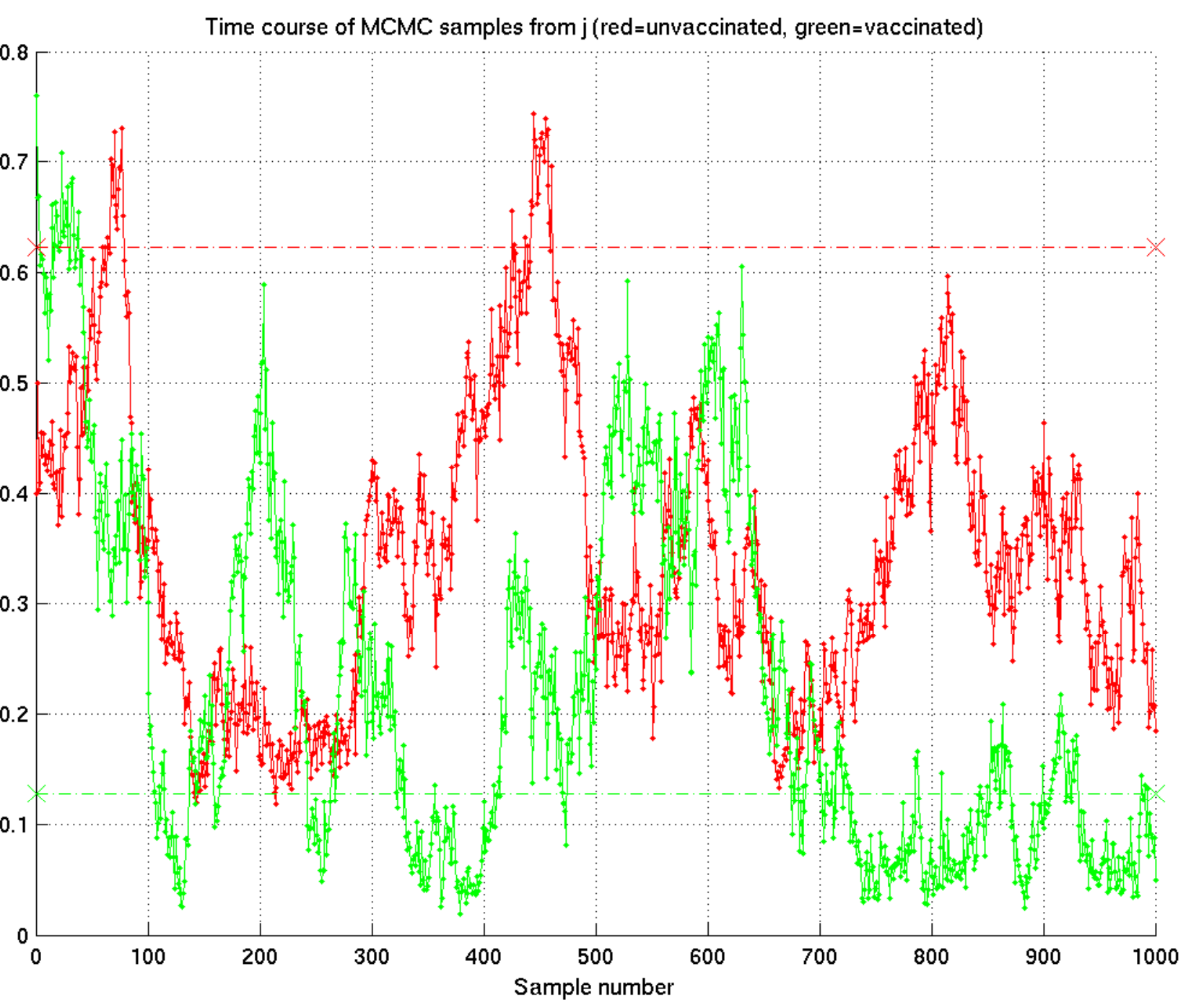}
\caption{The time-course of successive MCMC samples of $j$, showing
  how widely they vary, despite the relative constancy of the
  calculated near-observable probabilities in figure
  \ref{checkprobsfig}. The dotted lines indicate the true values.
\label{jfig}
}
\end{center}
\end{figure}

\begin{figure}[hpt]
\begin{center}
\includegraphics[scale=0.5]{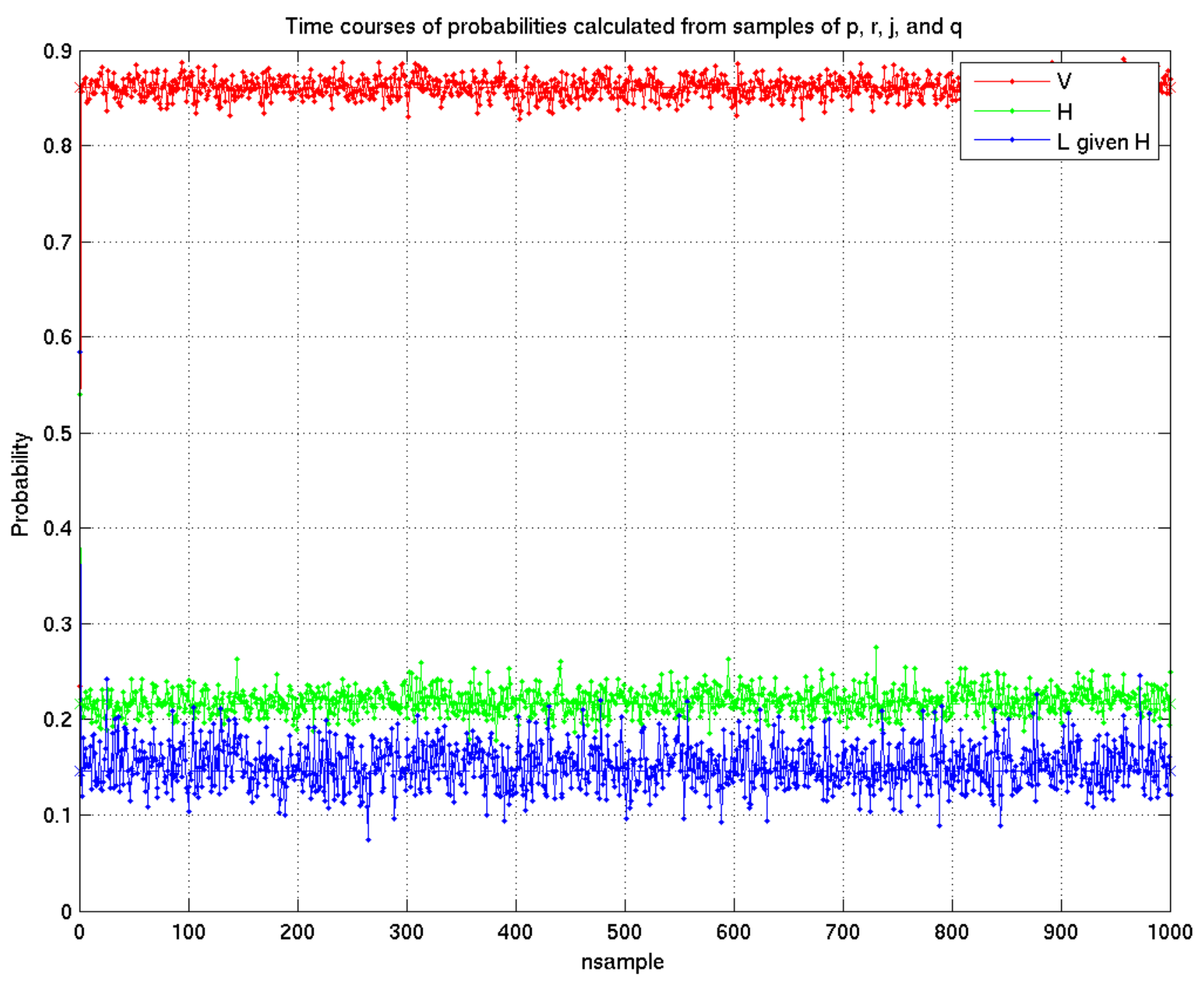}
\caption{The time-course of the probabilities of the observable
  variables showing how near constant they are, despite the widely
  varying sample probabilities in figure \ref{jfig}.
\label{checkprobsfig}
}
\end{center}
\end{figure}

First, to illustrate that samples of $j$ (and hence $T_0$) may vary
greatly without the fractions of vaccinated patients or hospitalised
patients in the population, or the fraction of covid-infected patients
in the hospitalised population, varying very much, we show the time
course of the sampled probabilities $j$ in figure \ref{jfig} and the
time course of the calculated probabilities of $V$, $H$, and $L|H$ in
figure \ref{checkprobsfig} (all from the same run from which we
consider specific samples below). This largely accounts for how it is
possible to have huge uncertainty in inferring $T_0$ which isn't
improved much by increasing the number of patients observed -- the
probabilities comprising $p,r,j,q$ can, with broad priors, easily vary
while keeping the probabilities of the observed data largely
unchanged.

Second, to check plausibility of the results in figure \ref{wideopen}
of section \ref{wideopensection} we give specific examples giving high
and low samples of $T_0$ from the posterior.

Specifically we look at the samples contributing to the specific
posterior distribution shown in figure \ref{wideopen} with
$T_{0,\text{true}}=-2.811$, where the 2.5th centile is at -7.804 and
the 97.5th centile at +0.284 . In this particular case the truth is
that vaccination is moderately protective, the crude method gives very
nearly the right result ($T_2=-2.68$), and TNCC is very
over-optimistic with $T_1=-4.81$.

The values of the various probabilities giving the maximum sample of
$T_0$ are shown in table \ref{samplevalsmax} and those giving the
minimum sample of $T_0$ are shown in table \ref{samplevalsmin}; in
both cases a verbal description and possible reasons for the values
seen are in the caption.

\begin{table}[hpt]
\begin{verbatim}
T0true is -2.811
  
Maximum T0 is at sample 11 and is 1.548
p = 0.878
r = 0.668, 0.896
j = 0.386, 0.747
q:
 for s=0:
       v=0    v=1
  l=0: 0.070, 0.704
  l=1: 0.054, 0.007
 for s=1:
       v=0    v=1
  l=0: 0.008, 0.782
  l=1: 0.364, 0.018
\end{verbatim}
\caption{Details of the sample contributing to the run in figure
  \ref{wideopen} with $T_{0,\text{true}}=-2.811$ and the maximum
  sample of $T_0$ from the posterior. In this case we have $j_V>j_v$
  indicating that vaccination makes it \textit{more} likely you will
  get covid (hence the high $T_0>0$). The values of $q$ say that in
  both groups people with non-covid illness are more likely to be
  hospitalised if they are vaccinated than if unvaccinated; this could
  be because of vaccine-induced injury, or because the vaccinated tend
  to seek healthcare more than the unvaccinated. In contrast those
  with covid are less likely to be hospitalised if vaccinated, perhaps
  because vaccination protects against hospitalisation but not against
  infection.
\label{samplevalsmax}
}
\end{table}

\begin{table}[hpt]
\begin{verbatim}
T0true is -2.811

Minimum T0 is at sample 492 and is -10.832
p = 0.946
r = 0.573, 0.883
j = 0.999, 0.017
q:
 for s=0:
       v=0    v=1
  l=0: 0.009, 0.848
  l=1: 0.246, 0.561
 for s=1:
       v=0    v=1
  l=0: 0.853, 0.187
  l=1: 0.168, 0.892
\end{verbatim}
\caption{Details of the sample contributing to the run in figure
  \ref{wideopen} with $T_{0,\text{true}}=-2.811$ and the minimum
  sample of $T_0$ from the posterior. In this case we have $j_V<j_v$,
  indicating that vaccination protects against infection. Most
  patients are in group 1 (as $p=0.946$). In group 1 the values of $q$
  indicate that patients with covid are \textit{more} likely to be
  hospitalised if vaccinated, perhaps because of vaccine-induced
  disease enhancement, while patients without covid are less likely to
  be admitted if vaccinated, perhaps because they are less worried
  by the possibility that they might get covid.
\label{samplevalsmin}
}
\end{table}

\clearpage 

\section{Appendix -- Values of $E$ corresponding to figures \ref{wideopen} --
  \ref{var3} and table \ref{LBresults}}
\label{Evalsappx}

This section lists the values of the traditional vaccine effectiveness
values as percentages for the truth $E_\text{true}$, the TNCC
estimate $E_1$, the crude estimate $E_2$, and the Bayesian 2.5th and
97.5th posterior centiles. In each table the runs are sorted by the
true value of $T_0$ in order to be able to relate them to the
corresponding items on the plots. The first two columns of each table
give the true value of $T_0$ and the simulated value of $P(L|v)$ for
that particular run.

Note, however, that the true value of $P(L|v)$ is not available to the
estimates, so that $E_1$ uses the estimate $\frac{\#LHv}{\#LHv +
  \#lHv}$, $E_2$ uses the estimate $\frac{\#LHv}{1 + \#v}$, and the
Bayesian centiles come from directly calculated values of $E_0$ for
each sample.

\begin{table}[hpt]
\begin{center}
\begin{tabular}{c|c|c|c|c|c|c|r}
$T_{0,\text{true}}$ & $P(L|v)_\text{true}$ & $E_\text{true}$ & $E_1$ & $E_2$ & Bayes $2.5$\% & Bayes $97.5$\% & $N_{\text{hosp}}$\\\hline
$-3.074$ & $0.710$ & $85.679$ & $87.073$ & $82.130$ & $56.083$ & $89.522$ & $776$\\\hline
$-2.811$ & $0.953$ & $42.287$ & $93.590$ & $91.957$ & $-24.365$ & $98.529$ & $216$\\\hline
$-2.733$ & $0.852$ & $68.037$ & $72.487$ & $49.826$ & $3.245$ & $79.283$ & $328$\\\hline
$-2.417$ & $0.623$ & $79.397$ & $86.691$ & $71.249$ & $-137.775$ & $90.977$ & $228$\\\hline
$-2.089$ & $0.750$ & $63.931$ & $74.134$ & $72.411$ & $42.955$ & $78.476$ & $682$\\\hline
$-1.110$ & $0.373$ & $56.086$ & $-969.141$ & $-1208.564$ & $-3663.482$ & $78.842$ & $189$\\\hline
$-0.332$ & $0.811$ & $6.920$ & $16.220$ & $47.658$ & $-81.462$ & $48.599$ & $382$\\\hline
$-0.312$ & $0.372$ & $18.680$ & $-23.207$ & $-52.304$ & $-147.062$ & $58.318$ & $470$\\\hline
$-0.144$ & $0.832$ & $2.534$ & $21.991$ & $40.760$ & $-72.299$ & $71.055$ & $404$\\\hline
$0.140$ & $0.318$ & $-9.806$ & $-25.147$ & $29.165$ & $-278.796$ & $41.253$ & $509$\\\hline
$0.141$ & $0.209$ & $-11.576$ & $-124.738$ & $-81.882$ & $-408.448$ & $59.018$ & $566$\\\hline
$0.288$ & $0.764$ & $-6.288$ & $-12.918$ & $-15.516$ & $-268.336$ & $31.548$ & $310$\\\hline
$0.472$ & $0.084$ & $-52.644$ & $-4.015$ & $-66.959$ & $-312.131$ & $77.324$ & $615$\\\hline
$0.709$ & $0.629$ & $-23.231$ & $-103.853$ & $-153.800$ & $-397.517$ & $18.837$ & $503$\\\hline
$1.084$ & $0.172$ & $-121.368$ & $-18.078$ & $-73.009$ & $-82.810$ & $34.445$ & $844$\\\hline
$1.249$ & $0.182$ & $-140.176$ & $-145.271$ & $-338.073$ & $-245.874$ & $18.932$ & $531$\\\hline
$1.279$ & $0.117$ & $-175.917$ & $-223.156$ & $-437.778$ & $-1244.194$ & $56.445$ & $556$\\\hline
$1.332$ & $0.153$ & $-165.567$ & $-3138.272$ & $-2473.715$ & $-8426.283$ & $-106.109$ & $768$\\\hline
$2.387$ & $0.195$ & $-271.490$ & $-255.245$ & $-186.354$ & $-388.993$ & $-43.838$ & $649$\\\hline
$3.613$ & $0.176$ & $-403.744$ & $-256.909$ & $-70.470$ & $-732.944$ & $-82.719$ & $447$\\\hline
\end{tabular}
\caption{Vaccine effectiveness ($E$, \%) values for each run shown in figure \ref{wideopen} 
corresponding to truth, TNCC, crude estimate, and the Bayesian 2.5th and 97.5th
centiles of the posterior for $T_0$.}
\end{center}
\end{table}

\begin{table}[hpt]
\begin{center}
\begin{tabular}{c|c|c|c|c|c|c|r}
$T_{0,\text{true}}$ & $P(L|v)_\text{true}$ & $E_\text{true}$ & $E_1$ & $E_2$ & Bayes $2.5$\% & Bayes $97.5$\% & $N_{\text{hosp}}$\\\hline
$-5.323$ & $0.840$ & $97.032$ & $95.640$ & $93.629$ & $37.582$ & $97.316$ & $4320$\\\hline
$-4.694$ & $0.984$ & $63.263$ & $72.943$ & $70.861$ & $32.001$ & $85.932$ & $4303$\\\hline
$-2.929$ & $0.913$ & $60.591$ & $58.247$ & $63.367$ & $10.578$ & $73.863$ & $5694$\\\hline
$-2.648$ & $0.475$ & $87.330$ & $85.178$ & $68.016$ & $-126.789$ & $93.423$ & $3696$\\\hline
$-2.414$ & $0.676$ & $76.727$ & $80.734$ & $68.697$ & $64.482$ & $83.411$ & $8119$\\\hline
$-2.285$ & $0.852$ & $56.605$ & $77.974$ & $74.986$ & $36.707$ & $86.546$ & $5229$\\\hline
$-2.136$ & $0.877$ & $47.786$ & $67.563$ & $32.371$ & $26.499$ & $73.487$ & $4708$\\\hline
$-1.859$ & $0.471$ & $74.158$ & $80.600$ & $79.880$ & $-6.115$ & $88.023$ & $6961$\\\hline
$-0.938$ & $0.418$ & $47.490$ & $4.737$ & $-100.291$ & $-268.590$ & $55.145$ & $6419$\\\hline
$-0.487$ & $0.212$ & $33.069$ & $-586.910$ & $-545.671$ & $-1568.546$ & $-6.846$ & $4394$\\\hline
$-0.032$ & $0.065$ & $2.969$ & $-15.644$ & $54.399$ & $-363.605$ & $77.810$ & $4978$\\\hline
$0.452$ & $0.424$ & $-26.464$ & $-7.999$ & $34.606$ & $-73.243$ & $36.370$ & $7202$\\\hline
$0.507$ & $0.226$ & $-44.476$ & $-10.675$ & $82.473$ & $-194.850$ & $82.507$ & $5673$\\\hline
$0.709$ & $0.629$ & $-23.231$ & $-96.972$ & $-155.118$ & $-238.697$ & $5.643$ & $5272$\\\hline
$1.416$ & $0.506$ & $-59.808$ & $-184.106$ & $-178.019$ & $-231.968$ & $-77.378$ & $6875$\\\hline
$1.617$ & $0.578$ & $-51.119$ & $-46.292$ & $50.894$ & $-123.061$ & $68.524$ & $4173$\\\hline
$1.955$ & $0.319$ & $-140.780$ & $-72.071$ & $-624.186$ & $-981.407$ & $-0.144$ & $3203$\\\hline
$2.050$ & $0.323$ & $-143.599$ & $-31.527$ & $-56.236$ & $-177.652$ & $19.011$ & $4550$\\\hline
$2.368$ & $0.066$ & $-552.585$ & $-198.120$ & $-304.273$ & $-821.188$ & $46.428$ & $4877$\\\hline
$5.182$ & $0.016$ & $-4654.035$ & $-3187.130$ & $-6954.207$ & $-8025.957$ & $-178.579$ & $5348$\\\hline
\end{tabular}
\caption{Vaccine effectiveness ($E$, \%) values for each run shown in figure \ref{wideopenmore} 
corresponding to truth, TNCC, crude estimate, and the Bayesian 2.5th and 97.5th
centiles of the posterior for $T_0$.}
\end{center}
\end{table}

\begin{table}[hpt]
\begin{center}
\begin{tabular}{c|c|c|c|c|c|c|r}
$T_{0,\text{true}}$ & $P(L|v)_\text{true}$ & $E_\text{true}$ & $E_1$ & $E_2$ & Bayes $2.5$\% & Bayes $97.5$\% & $N_{\text{hosp}}$\\\hline
$-3.241$ & $0.775$ & $84.677$ & $86.052$ & $90.246$ & $50.718$ & $91.565$ & $632$\\\hline
$-3.102$ & $0.887$ & $70.565$ & $73.177$ & $17.053$ & $53.642$ & $76.278$ & $768$\\\hline
$-2.290$ & $0.323$ & $85.721$ & $77.985$ & $81.421$ & $-245.818$ & $91.606$ & $273$\\\hline
$-2.068$ & $0.399$ & $80.578$ & $91.767$ & $84.117$ & $42.121$ & $94.678$ & $922$\\\hline
$-1.836$ & $0.699$ & $61.383$ & $3.424$ & $-86.894$ & $-214.846$ & $94.521$ & $369$\\\hline
$-1.717$ & $0.960$ & $15.416$ & $6.478$ & $-718.851$ & $-220.415$ & $55.120$ & $119$\\\hline
$-1.444$ & $0.419$ & $65.287$ & $47.412$ & $31.686$ & $-18.482$ & $78.275$ & $466$\\\hline
$-1.167$ & $0.413$ & $56.491$ & $47.637$ & $50.346$ & $-17.220$ & $62.671$ & $794$\\\hline
$-0.838$ & $0.887$ & $12.870$ & $-83.351$ & $-820.894$ & $-1033.070$ & $7.428$ & $679$\\\hline
$-0.044$ & $0.233$ & $3.359$ & $-209.900$ & $-98.746$ & $-446.878$ & $39.824$ & $558$\\\hline
$0.046$ & $0.591$ & $-1.886$ & $22.118$ & $36.131$ & $-6.122$ & $57.717$ & $608$\\\hline
$0.187$ & $0.733$ & $-4.771$ & $11.021$ & $75.733$ & $-15.422$ & $77.879$ & $200$\\\hline
$0.615$ & $0.514$ & $-28.711$ & $-95.077$ & $-163.170$ & $-164.313$ & $-20.107$ & $648$\\\hline
$1.016$ & $0.148$ & $-119.117$ & $-81.868$ & $72.837$ & $-585.205$ & $91.567$ & $666$\\\hline
$1.452$ & $0.234$ & $-141.960$ & $83.206$ & $95.855$ & $-4100.312$ & $36.774$ & $262$\\\hline
$1.727$ & $0.029$ & $-396.603$ & $-991.667$ & $-1221.273$ & $-1608.374$ & $91.670$ & $235$\\\hline
$1.748$ & $0.063$ & $-341.544$ & $-342.359$ & $-547.313$ & $-1192.383$ & $48.411$ & $580$\\\hline
$2.267$ & $0.684$ & $-39.463$ & $-8.218$ & $-57.621$ & $-79.587$ & $1.403$ & $946$\\\hline
$4.107$ & $0.012$ & $-3492.194$ & $-8227.568$ & $-1862.944$ & $-20366.003$ & $-1728.672$ & $519$\\\hline
$5.511$ & $0.066$ & $-1326.968$ & $-1237.600$ & $-74.263$ & $-1003.423$ & $79.109$ & $467$\\\hline
\end{tabular}
\caption{Vaccine effectiveness ($E$, \%) values for each run shown in figure \ref{var2} 
corresponding to truth, TNCC, crude estimate, and the Bayesian 2.5th and 97.5th
centiles of the posterior for $T_0$.}
\end{center}
\end{table}

\begin{table}[hpt]
\begin{center}
\begin{tabular}{c|c|c|c|c|c|c|r}
$T_{0,\text{true}}$ & $P(L|v)_\text{true}$ & $E_\text{true}$ & $E_1$ & $E_2$ & Bayes $2.5$\% & Bayes $97.5$\% & $N_{\text{hosp}}$\\\hline
$-7.010$ & $0.996$ & $83.049$ & $81.166$ & $57.440$ & $77.896$ & $84.463$ & $677$\\\hline
$-2.446$ & $0.932$ & $41.659$ & $46.092$ & $-38.647$ & $41.406$ & $50.690$ & $713$\\\hline
$-1.884$ & $0.724$ & $60.600$ & $64.722$ & $23.505$ & $58.465$ & $70.133$ & $703$\\\hline
$-1.533$ & $0.109$ & $76.382$ & $75.208$ & $39.189$ & $57.085$ & $86.903$ & $674$\\\hline
$-0.891$ & $0.495$ & $42.040$ & $48.369$ & $-27.738$ & $37.268$ & $57.713$ & $662$\\\hline
$-0.861$ & $0.472$ & $41.908$ & $23.387$ & $-80.493$ & $2.731$ & $40.895$ & $736$\\\hline
$-0.323$ & $0.746$ & $8.824$ & $0.078$ & $-145.494$ & $-13.428$ & $11.202$ & $697$\\\hline
$-0.282$ & $0.423$ & $15.824$ & $17.889$ & $-122.686$ & $-6.651$ & $34.081$ & $695$\\\hline
$-0.072$ & $0.211$ & $5.578$ & $-8.134$ & $-175.723$ & $-66.488$ & $22.612$ & $701$\\\hline
$0.148$ & $0.053$ & $-15.007$ & $60.607$ & $-11.625$ & $19.171$ & $77.653$ & $662$\\\hline
$0.281$ & $0.748$ & $-6.588$ & $-9.331$ & $-170.250$ & $-20.523$ & $-0.381$ & $695$\\\hline
$0.296$ & $0.565$ & $-12.579$ & $-5.699$ & $-128.836$ & $-28.103$ & $8.668$ & $741$\\\hline
$0.353$ & $0.597$ & $-13.596$ & $-6.247$ & $-175.058$ & $-26.495$ & $8.581$ & $715$\\\hline
$0.461$ & $0.231$ & $-39.591$ & $-18.828$ & $-192.822$ & $-63.476$ & $11.420$ & $703$\\\hline
$1.453$ & $0.477$ & $-66.950$ & $-72.127$ & $-327.003$ & $-104.348$ & $-43.129$ & $680$\\\hline
$1.567$ & $0.333$ & $-111.813$ & $-150.507$ & $-445.982$ & $-225.309$ & $-96.712$ & $672$\\\hline
$1.591$ & $0.692$ & $-32.450$ & $-22.656$ & $-189.307$ & $-36.365$ & $-12.549$ & $712$\\\hline
$2.132$ & $0.158$ & $-287.524$ & $-654.600$ & $-1834.200$ & $-1450.494$ & $-359.874$ & $753$\\\hline
$2.923$ & $0.236$ & $-260.594$ & $-306.154$ & $-970.408$ & $-494.422$ & $-207.465$ & $693$\\\hline
$4.623$ & $0.032$ & $-2329.143$ & $-1613.424$ & $-4356.143$ & $-3817.201$ & $-858.843$ & $651$\\\hline
\end{tabular}
\caption{Vaccine effectiveness ($E$, \%) values for each run shown in figure \ref{var1} 
corresponding to truth, TNCC, crude estimate, and the Bayesian 2.5th and 97.5th
centiles of the posterior for $T_0$.}
\end{center}
\end{table}

\begin{table}[hpt]
\begin{center}
\begin{tabular}{c|c|c|c|c|c|c|r}
$T_{0,\text{true}}$ & $P(L|v)_\text{true}$ & $E_\text{true}$ & $E_1$ & $E_2$ & Bayes $2.5$\% & Bayes $97.5$\% & $N_{\text{hosp}}$\\\hline
$-3.732$ & $0.634$ & $93.722$ & $94.079$ & $84.328$ & $93.151$ & $95.133$ & $6520$\\\hline
$-2.529$ & $0.309$ & $88.858$ & $89.610$ & $75.642$ & $87.741$ & $91.733$ & $7253$\\\hline
$-2.484$ & $0.402$ & $86.787$ & $88.381$ & $70.316$ & $85.010$ & $89.951$ & $6880$\\\hline
$-2.239$ & $0.993$ & $5.709$ & $6.649$ & $-120.503$ & $5.750$ & $7.548$ & $6741$\\\hline
$-1.307$ & $0.451$ & $59.665$ & $62.298$ & $8.423$ & $56.807$ & $67.505$ & $6953$\\\hline
$-1.278$ & $0.304$ & $64.314$ & $64.383$ & $15.461$ & $58.471$ & $71.697$ & $7040$\\\hline
$-1.252$ & $0.908$ & $18.606$ & $18.278$ & $-98.120$ & $15.488$ & $20.346$ & $6771$\\\hline
$-0.999$ & $0.363$ & $52.214$ & $49.184$ & $-27.765$ & $38.659$ & $56.818$ & $7141$\\\hline
$-0.982$ & $0.423$ & $49.044$ & $48.812$ & $-26.332$ & $39.327$ & $56.453$ & $7362$\\\hline
$-0.883$ & $0.533$ & $39.844$ & $41.428$ & $-61.820$ & $33.721$ & $47.559$ & $6732$\\\hline
$-0.677$ & $0.658$ & $24.886$ & $29.810$ & $-94.370$ & $25.507$ & $35.262$ & $7056$\\\hline
$-0.565$ & $0.854$ & $9.993$ & $10.366$ & $-108.529$ & $5.924$ & $12.382$ & $6822$\\\hline
$-0.169$ & $0.445$ & $9.277$ & $11.907$ & $-135.263$ & $-8.819$ & $24.917$ & $6769$\\\hline
$0.263$ & $0.735$ & $-6.529$ & $-12.397$ & $-184.982$ & $-21.206$ & $-5.665$ & $6995$\\\hline
$0.327$ & $0.844$ & $-4.545$ & $-4.249$ & $-143.990$ & $-9.198$ & $-0.827$ & $6649$\\\hline
$0.521$ & $0.809$ & $-8.407$ & $-9.639$ & $-175.911$ & $-15.457$ & $-4.683$ & $6532$\\\hline
$0.890$ & $0.019$ & $-137.030$ & $-64.789$ & $-324.022$ & $-141.645$ & $-23.710$ & $6761$\\\hline
$1.636$ & $0.662$ & $-37.328$ & $-39.671$ & $-248.139$ & $-53.268$ & $-31.535$ & $7148$\\\hline
$2.132$ & $0.158$ & $-287.524$ & $-299.566$ & $-908.024$ & $-401.116$ & $-219.834$ & $7497$\\\hline
$5.250$ & $0.711$ & $-40.359$ & $-34.858$ & $-248.474$ & $-41.094$ & $-26.099$ & $6787$\\\hline
\end{tabular}
\caption{Vaccine effectiveness ($E$, \%) values for each run shown in figure \ref{var1more} 
corresponding to truth, TNCC, crude estimate, and the Bayesian 2.5th and 97.5th
centiles of the posterior for $T_0$.}
\end{center}
\end{table}

\begin{table}[hpt]
\begin{center}
\begin{tabular}{c|c|c|c|c|c|c|r}
$T_{0,\text{true}}$ & $P(L|v)_\text{true}$ & $E_\text{true}$ & $E_1$ & $E_2$ & Bayes $2.5$\% & Bayes $97.5$\% & $N_{\text{hosp}}$\\\hline
$-3.376$ & $0.940$ & $62.853$ & $66.303$ & $64.809$ & $59.614$ & $68.364$ & $969$\\\hline
$-2.308$ & $0.822$ & $61.681$ & $66.420$ & $59.239$ & $53.676$ & $64.287$ & $906$\\\hline
$-1.072$ & $0.432$ & $52.202$ & $80.137$ & $56.796$ & $46.779$ & $65.238$ & $734$\\\hline
$-0.858$ & $0.890$ & $12.951$ & $20.380$ & $11.856$ & $6.066$ & $16.209$ & $952$\\\hline
$-0.778$ & $0.299$ & $45.192$ & $80.474$ & $44.283$ & $30.447$ & $56.173$ & $658$\\\hline
$0.520$ & $0.294$ & $-40.083$ & $52.804$ & $-24.963$ & $-52.152$ & $-6.494$ & $719$\\\hline
$0.546$ & $0.507$ & $-26.194$ & $23.196$ & $-39.530$ & $-55.869$ & $-23.624$ & $786$\\\hline
$0.720$ & $0.635$ & $-23.102$ & $12.756$ & $-26.887$ & $-36.252$ & $-16.690$ & $837$\\\hline
$1.404$ & $0.382$ & $-87.406$ & $8.868$ & $-99.850$ & $-124.598$ & $-75.480$ & $733$\\\hline
$1.461$ & $0.745$ & $-24.339$ & $5.409$ & $-26.526$ & $-33.467$ & $-19.013$ & $863$\\\hline
$1.651$ & $0.613$ & $-45.518$ & $4.762$ & $-43.508$ & $-55.745$ & $-34.639$ & $808$\\\hline
$1.726$ & $0.739$ & $-27.351$ & $5.127$ & $-29.414$ & $-37.998$ & $-21.800$ & $843$\\\hline
$1.900$ & $0.324$ & $-135.031$ & $0.098$ & $-183.903$ & $-234.442$ & $-141.947$ & $693$\\\hline
$2.161$ & $0.326$ & $-147.737$ & $-2.670$ & $-139.846$ & $-173.053$ & $-109.532$ & $736$\\\hline
$2.186$ & $0.280$ & $-177.197$ & $7.157$ & $-159.727$ & $-202.262$ & $-124.833$ & $691$\\\hline
$2.201$ & $0.447$ & $-96.730$ & $-0.500$ & $-101.393$ & $-126.738$ & $-81.903$ & $755$\\\hline
$2.352$ & $0.556$ & $-67.187$ & $-0.076$ & $-59.844$ & $-76.711$ & $-48.576$ & $820$\\\hline
$3.413$ & $0.100$ & $-669.652$ & $-26.932$ & $-581.962$ & $-762.468$ & $-444.741$ & $568$\\\hline
$3.561$ & $0.770$ & $-28.747$ & $-0.848$ & $-29.128$ & $-34.445$ & $-22.099$ & $884$\\\hline
$7.070$ & $0.651$ & $-53.545$ & $-4.471$ & $-58.627$ & $-71.408$ & $-48.897$ & $835$\\\hline
\end{tabular}
\caption{Vaccine effectiveness ($E$, \%) values for each run shown in figure \ref{var3} 
corresponding to truth, TNCC, crude estimate, and the Bayesian 2.5th and 97.5th
centiles of the posterior for $T_0$.}
\end{center}
\end{table}

\begin{table}[hpt]
\begin{center}
\begin{tabular}{c|c|c|c|c|c}
& \multicolumn{4}{c|}{Vaccine effectiveness (percentages)} & \\
Assumption & $E_1$ & $E_2$ & Bayes $2.5\%$ & Bayes $97.5\%$ & $N_{\text{hosp}}$\\\hline
1 & 91.5 & 91.5 & -121 & +96.8 & 127820\\\hline
2 & 91.5 & 99.3 & -902 & +99.1 & 127820\\\hline
3 & 91.5 & 91.5 & -48.2 & +98.9 & 127820\\\hline
\end{tabular}
\caption{Vaccine effectiveness ($E,\%$) values using real-life data
  as in table \ref{LB} and the three assumptions listed in section
  \ref{realdataintro}.
\label{LBresultsE}
}
\end{center}
\end{table}

\clearpage

\bibliography{ms}

\begin{thebibliography}{10}

\bibitem{Chua}
{\relax Huiying Chua, Shuo Feng, et. al.}, ``\relax{The use of test-negative
  controls to monitor vaccine effectiveness: a systematic review of
  methodology},'' {\em Epidemiology}, vol.~31, no.~1, pp.~43--64, 2020.

\bibitem{KirsebomLancet}
F.~C.~M. Kirsebom, N.~Andrews, J.~Stowe, S.~Toffa, R.~Sachdeva, E.~Gallagher,
  N.~Groves, A.-M. O'Connell, M.~Chand, M.~Ramsay, and J.~L. Bernal, ``Covid-19
  vaccine effectiveness against the omicron (ba.2) variant in england,'' {\em
  The Lancet Infectious Diseases}, vol.~22, no.~7, pp.~931--933, 2022.

\bibitem{Week48}
{\relax U.K Health Security Agency}, ``{\relax COVID-19 vaccine surveillance
  report - Week 48}.''
  \url{https://assets.publishing.service.gov.uk/government/uploads/system/uploads/attachment_data/file/1037987/Vaccine-surveillance-report-week-48.pdf},
  2021.
\newblock Retrieved on 28.6.2022.

\bibitem{Stowe}
{\relax Julia Stowe, Nick Andrews, Freja Kirsebom, Mary Ramsay, Jamie Lopez
  Bernal }, ``\relax{ Effectiveness of COVID-19 vaccines against Omicron and
  Delta hospitalisation: test negative case-control study }.''
  \url{https://www.medrxiv.org/content/10.1101/2022.04.01.22273281v1.full-text},
  2022.
\newblock Retrieved on 28.6.2022.

\bibitem{Broome}
\relax{Broome CV, Facklam RR, Fraser DW}, ``\relax{Pneumococcal disease after
  pneumococcal vaccination: an alternative method to estimate the efficacy of
  pneumococcal vaccine},'' {\em New England Journal of Medicine}, vol.~303,
  no.~10, pp.~549--552, 1980.

\bibitem{Sullivan}
{\relax Sheena G. Sullivan, Eric J. Tchetgen Tchetgen, and Benjamin J.
  Cowling}, ``{\relax Theoretical Basis of the Test-Negative Study Design for
  Assessment of Influenza Vaccine Effectiveness},'' {\em American Journal of
  Epidemiology}, vol.~184, no.~5, pp.~345--353, 2016.

\bibitem{Westreich}
{\relax Daniel Westreich and Michael G. Hudgens}, ``{\relax Beware the
  Test-Negative Design},'' {\em American Journal of Epidemiology}, vol.~184,
  no.~5, pp.~354--356, 2016.

\bibitem{LopezBernalNEJM}
{\relax Jamie Lopez Bernal et. al.}, ``{\relax Effectiveness of Covid-19
  Vaccines against the B.1.617.2 (Delta) Variant},'' {\em New England Journal
  of Medicine}, vol.~385, no.~7, pp.~585--594, 2021.

\bibitem{GBnews}
{\relax GB news channel}.
  \url{https://twitter.com/resist_05/status/1511895388185571328}, 2022.
\newblock Retrieved on 04.7.2022.

\bibitem{RCTreview}
M.~L. Meldrum, ``A brief history of the randomized controlled trial,'' {\em
  Haematology/Oncology clinics of North America}, vol.~14, pp.~745--760, Aug.
  2000.

\bibitem{Pfizer1}
{\relax Edward E. Walsh et. al.}, ``{\relax Safety and Immunogenicity of Two
  RNA-Based Covid-19 Vaccine Candidates},'' {\em New England Journal of
  Medicine}, vol.~383, no.~25, pp.~2439--2450, 2020.

\bibitem{Pfizer2}
{\relax Fernando P. Polack et. al.}, ``{\relax Safety and Efficacy of the
  BNT162b2 mRNA Covid-19 Vaccine},'' {\em New England Journal of Medicine},
  vol.~383, no.~27, pp.~2603--2615, 2020.

\bibitem{Pfizer3}
{\relax S.J. Thomas et. al.}, ``{\relax Safety and Efficacy of the BNT162b2
  mRNA Covid-19 Vaccine through 6 Months},'' {\em New England Journal of
  Medicine}, vol.~385, no.~19, pp.~1761--1773, 2021.

\bibitem{AZvaccine}
{\relax Merryn Voysey et. al.}, ``{\relax Safety and efficacy of the ChAdOx1
  nCoV-19 vaccine (AZD1222) against SARS-CoV-2: an interim analysis of four
  randomised controlled trials in Brazil, South Africa, and the UK},'' {\em
  Lancet}, vol.~397, pp.~99--111, 2021.

\bibitem{Doshi}
{\relax Joseph Fraiman, Juan Erviti, Mark Jones, Sander Greenland, Patrick
  Whelan, Robert M. Kaplan, Peter Doshi}, ``{\relax Serious adverse events of
  special interest following mRNA vaccination in randomized trials}.''
  \url{https://ssrn.com/abstract=4125239}, 2022.
\newblock Retrieved on 30.6.2022.

\bibitem{Ventavia}
P.~D. Thacker, ``{\relax Covid-19: Researcher blows the whistle on data
  integrity issues in Pfizer's vaccine trial},'' {\em BMJ}, vol.~375, 2021.

\bibitem{rnealMCMC}
{\relax Radford M. Neal}, ``{\relax Probabilistic inference using Markov Chain
  Monte Carlo Methods},'' {\em 1993}, Technical Report CRG-TR-93-1, Department
  of Computer Science, University of Toronto.

\bibitem{Dagpunar}
{\relax John Dagpunar}, {\em {\relax Principles of Random Variate Generation
  }}.
\newblock {\relax Clarendon Press}, 1988.

\bibitem{WhitworthLifetime}
{\relax Laura Whitworth et. al.}, ``\relax{A Bayesian analysis of the
  association between Leukotriene A4 Hydrolase genotype and survival in
  tuberculous meningitis},'' {\em eLife}, vol.~10, 2021.
\newblock {\relax Downloaded from https://elifesciences.org/articles/61722 on
  31.1.2021}.

\end{thebibliography}
\bibliographystyle{ieeetr}

\end{document}